\documentclass[aps,prb,reprint,showpacs,floatfix]{revtex4-2}

\usepackage{longtable}
\usepackage{xurl}
\AtBeginDocument{\renewcommand{\eprint}[2][]{\url{#2}}}
\usepackage[final]{graphicx}
\usepackage{capt-of}
\usepackage{dcolumn}
\usepackage{bm}
\usepackage{braket}
\usepackage{amsmath}

\usepackage{xcolor}

\newcommand{\mm}[1]     {\ifmmode {#1} \else{}${#1}$\fi}
\newcommand{\mmm}[1]    {\ifmmode{}#1 \else{}${#1}$\fi}
\newcommand{\beq}[1]{\begin{equation}\label{#1}}
\newcommand{\eeq}{\end{equation}}

\newcommand{\beqm}[1]{\begin{multiline}\label{#1}}
\newcommand{\eeqm}{\end{multiline}}

\def\P6m{\mm{P{6}/m}}

\begin{document}

\title{\large Crystal and magnetic structures of MnSi in zero magnetic field: Neutron diffraction and symmetry analysis}

\author{V.~Pomjakushin}
\affiliation{Laboratory for Neutron Scattering and Imaging, Paul Scherrer
Institut, CH-5232 Villigen PSI, Switzerland
}
\author{I.~Plokhikh}
\affiliation{Laboratory for Multiscale Materials Experiments, Paul Scherrer Institut, CH$-$5232 Villigen PSI, Switzerland
}
\affiliation{Physics Department, Technical University Dortmund, Otto-Hahn-Stra{\ss}e 4, 44227 Dortmund, Germany}
\affiliation{Research Center Future Energy Materials and Systems, 44227 Dortmund, Germany}
\author{ D.~H.~Fabini}
\affiliation{Laboratory for Neutron Scattering and Imaging, Paul Scherrer
Institut, CH$-$5232 Villigen PSI, Switzerland
}

\author{E.~Pomjakushina}
\affiliation{Laboratory for Multiscale Materials Experiments, Paul Scherrer Institut, CH-5232 Villigen PSI, Switzerland}

\date{November 7, 2026}

\begin{abstract}
Combining high-resolution wide-angle neutron powder diffraction with magnetic-superspace symmetry analysis, we refine the crystal structure and present, to our knowledge, the first systematic refinement of cubic $P2_1 3$ MnSi's zero-field magnetic structure across multiple magnetic satellite families. We compare single-arm rhombohedral  $3+1$D and cubic four-arm $3+4$D magnetic superspace-group (MSSG) descriptions with propagation vectors along $\langle111\rangle$. Simultaneous extinction of the two satellite groups around $\{100\}$ dictates a transverse circular-helix model. This suppression and refinements of 18 separately varied satellite-group intensities support this ideal limit with only one adjustable magnetic amplitude. In the ideal single-arm model with MSSG $R3.1'(0,0,g)ts$, the moment has constant magnitude $0.535(3)\,\mu_{\rm B}$ per Mn. An explicit four-arm $3+4$D cubic superspace construction presented here combines four interpenetrating helices and, for the illustrated phase choice, has moments ranging from zero to $0.874(5)\,\mu_{\rm B}$ per Mn, with the same root-mean-square moment as the $3+1$D structure. Their identical powder profiles prevent the present measurements from distinguishing them. The average crystal structure remains cubic within experimental resolution, providing structural motivation for the cubic magnetic model. The mean lattice parameter, determined with formal relative precision $\sigma_a/a\lesssim10^{-5}$, exhibits nonmonotonic temperature dependence below the N\'eel temperature.
\end{abstract}

\maketitle

\section{Introduction}

MnSi is a prototypical itinerant chiral magnet with the noncentrosymmetric cubic B20 structure, space group $P2_1 3$. Competition between predominantly ferromagnetic exchange and the Dzyaloshinskii--Moriya interaction favors a long-wavelength modulation, while weak anisotropy selects the $\langle111\rangle$ propagation directions~\cite{BakJensen1980}. At ambient pressure, magnetic order develops near $T_N\simeq29$~K through a weakly first-order transition~\cite{Stishov2007}. Its field-induced skyrmion lattice has made MnSi a central example of topological magnetism~\cite{doi:10.1126/science.1166767}, but its zero-field structure nevertheless merits separate examination.

Transport measurements provide complementary evidence for topological spin textures. A topological Hall effect (THE) was reported in the finite-field $A$ phase of MnSi close to $T_N$~\cite{Neubauer2009}. In the isostructural short-period helimagnet MnGe, a substantially larger Hall contribution attributed to THE occurs over a broader temperature range~\cite{kanazawa2021}. Our zero-field diffraction analysis and the continuum topology discussed below do not establish a connection between the proposed four-arm structure and THE.

Early magnetization and neutron measurements already indicated that MnSi was not a simple ferromagnet~\cite{Levinson1973}. Ishikawa \textit{et al.} subsequently observed satellites near $(110)$ and $(111)$ using triple-axis diffraction with $\lambda=2.55$~\AA, and around the reciprocal-space origin using small-angle neutron scattering (SANS) with $\lambda=4.99$~\AA~\cite{Ishikawa1976}. They established propagation along $\langle111\rangle$ with a period of approximately 180~\AA\ at 4.2~K. Their zero-field diffraction did not, however, determine the spin-rotation plane: a plane perpendicular to the propagation direction was inferred with support from the magnetization. The quoted moment of approximately $0.4\,\mu_B$ per Mn likewise came from magnetization. Thus, the helical interpretation was experimentally well motivated, but was not a refinement of all site-resolved Fourier components against zero-field satellite intensities.

Subsequent polarized small-angle neutron diffraction established the relation between crystal chirality and magnetic helicity, including measurements at 16~K in a 60~Oe guide field after zero-field cooling~\cite{Ishida1985}. These low-field measurements constrain the helicity of the modulation, but helicity alone does not confirm a single-arm structure. At a resolved  magnetic satellite, the polarization dependence probes the Fourier component of the corresponding propagation-vector arm, without determining relative phases between distinct arms. The same helicity signature is therefore compatible with single-arm $3+1$D domains or with the corresponding arm of a coherent four-arm $3+4$D structure. Our study addresses these alternatives in zero applied magnetic field.

Polarized SANS further characterized domain selection and the field evolution of the helix~\cite{Grigoriev2006}, while symmetry analyses examined the coupling between structural and magnetic chirality~\cite{Dmitriev2012}. Complementary, symmetry-constrained muon spin rotation analyses proposed a small relative phase shift between inequivalent Mn orbits of the single-arm structure~\cite{DalmasDeReotier2016,DalmasDeReotier2018}. These results underline the distinction between establishing the long-period modulation and determining its detailed magnetic structure within the crystallographic unit cell. 

To our knowledge, the present work provides the first systematic refinement of the zero-field magnetic structure of MnSi from high-resolution wide-angle neutron powder diffraction intensities across multiple satellite families. Such a refinement requires resolving weak magnetic satellites beside much stronger nuclear reflections and reliably subtracting matched low- and high-temperature patterns. The analysed patterns span approximately $Q=0.5$--$3.8$~\AA$^{-1}$. The calculated reflection list contains 320 entries grouped into 20 distinct powder positions, providing quantitative intensity constraints on the symmetry-allowed magnetic models.

Our earlier neutron-diffraction and symmetry study of MnGe demonstrated the importance of explicitly comparing symmetry-allowed single- and multi-arm structures~\cite{PomjakushinMnGe2023}. Several models could describe the powder magnetic intensities, whereas the resolved orthorhombic structural distortion supplied a further symmetry constraint. MnSi poses a related question with a different propagation-vector star: four $\langle111\rangle$ arms, rather than the three orthogonal $\langle100\rangle$ arms considered for MnGe.

Here we combine magnetic-superspace symmetry with high-resolution powder diffraction to constrain the zero-field structure of MnSi. We compare a single-arm $3+1$D description with an explicit four-arm $3+4$D construction that retains the phase transformations of all four active arms, as detailed in Sec.~\ref{symmetry}. This approach also identifies a remaining limitation: the ideal-helix single-arm and four-arm models supported by the refinements yield identical powder profiles after the appropriate amplitude normalization.

\section{Sample synthesis and experimental methods}
\label{exp}

The polycrystalline MnSi sample was prepared by arc melting Mn and Si, each of at least 99.99\% purity, in a compact MAM-1 arc melter (Edmund B\"uhler GmbH)~\cite{BuehlerMAM1}. To improve homogeneity, the sample was flipped and remelted several times, then annealed at $1000\,{}^\circ$C for 48~h.

Neutron powder diffraction measurements were performed at the SINQ spallation source, Paul Scherrer Institute, Switzerland, using the high-resolution thermal-neutron diffractometer HRPT~\cite{hrpt}. Wavelengths of 1.155, 1.494 and approximately 2.45~\AA\ were used. The high-intensity (HI) and high-resolution (HR) modes employ primary white-beam collimations of $40'$, and $6'$, respectively, with relative resolutions $\delta d/d>1.8\times10^{-3}$, and $>0.9\times10^{-3}$~\cite{hrpt}. The HR intensity is approximately 7\% of the HI intensity.  HI mode was used for the weak magnetic diffraction, while HR mode provided greater sensitivity to fine details of the crystal structure.

\begin{figure*}[!t]
\centering
\includegraphics[width=\textwidth]{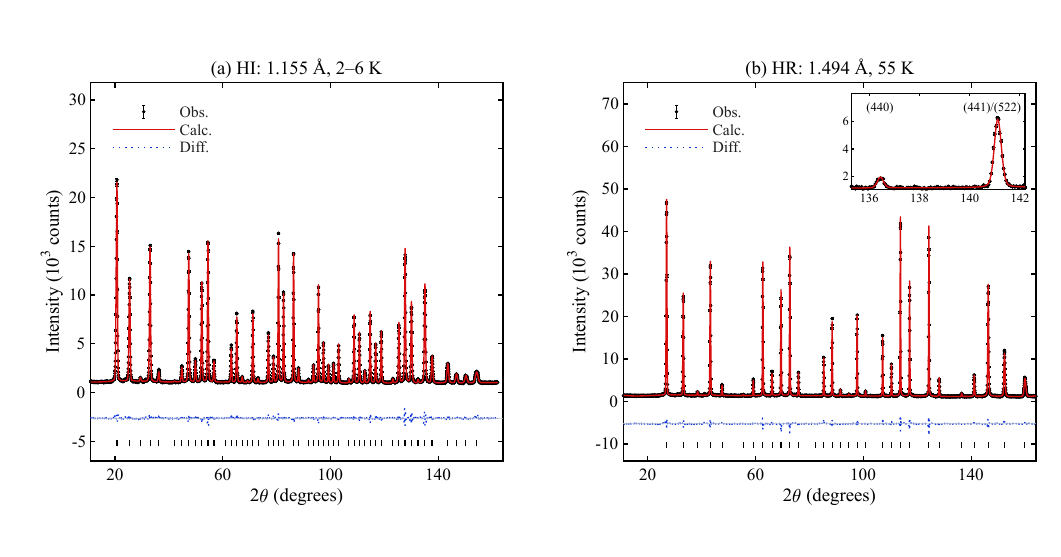}
\caption{Nuclear Rietveld refinements of MnSi in $P2_13$: (a) the pattern obtained by merging the 2--6~K datasets measured at HRPT with $\lambda=1.155$~\AA\ in HI mode, and (b) the 55~K pattern measured with $\lambda=1.494$~\AA\ in HR mode. Black symbols and vertical bars show the data points with standard uncertainties; many bars are smaller than the symbols. Red curves show the calculated profiles, blue dotted curves the observed-minus-calculated differences shifted downwards, and vertical ticks the Bragg positions. The inset in (b) shows the high-angle $(440)$ and $(441)/(522)$ reflections, with the same angular and intensity units as the main panel. The reliability factors are $R_{\rm wp}=4.32\%$, $\chi^2=3.928$ for (a), and $R_{\rm wp}=5.53\%$, $\chi^2=3.195$ for (b), using FullProf's  conventions.}
\label{dpatt}
\end{figure*}

The magnetic-structure analysis uses three independent HRPT datasets, labelled H1, H2 and H3, measured at $\lambda\simeq2.45$~\AA\ in HI mode. Each comprises a base-temperature 1.5–1.8 K pattern and a 54~K reference pattern. H3, used for the representative profile comparison, contains approximately 119 and 90~h of counting at base temperature and 54~K, respectively. The three base-minus-54~K difference patterns were analysed independently on their native grids with $0.05^\circ$ steps in $2\theta$, retaining each dataset's nuclear scale, angular calibration and peak-profile parameters.

The structural measurements at $\lambda=1.155$~\AA\ form a separate HRPT temperature-scan dataset comprising many individual diffraction patterns. For the low-temperature crystal-structure refinement, the datasets collected over 2--6~K were merged to improve counting statistics, yielding the diffraction pattern shown in Fig.~\ref{dpatt}(a). The lattice-parameter curve in Fig.~\ref{c_of_T} is compiled from 110 independently refined patterns, with the individual results and their averages in 1~K intervals shown together. 

Complementary measurements were performed on the cold-neutron powder diffractometer DMC at $\lambda\simeq2.45$~\AA. A base-minus-54~K dataset, on its native $0.125^\circ$ grid in $2\theta$, provides a high-statistics check of the absence of the first $\{100\}$ satellites. DMC was used for the magnetic temperature scans because its full-detector count rates estimated from MnSi nuclear peaks were approximately 50--60 times those of HRPT in HI mode, allowing useful statistics in much shorter reasonable counting intervals. The quantitative magnetic profile fits are based on HRPT datasets. Experimental details, DMC profile diagnostics and subtraction limitations, and crystal-structure checks are provided in Secs.~S1--S3 of the Supplementary Material (SM)~\cite{SupplementalMaterial}. Hereafter, S-prefixed section, figure and table numbers refer to the SM.

FullProf~\cite{Fullprof} was used for crystal- and magnetic-structure refinements and for calculating diffraction profiles, using its tabulated neutron scattering lengths. The complementary weighted profile-decomposition method is described in Sec.~\ref{sec:lebail}. Symmetry analysis was performed using ISODISTORT from the ISOTROPY software suite~\cite{isod,isod2} and tools of the Bilbao Crystallographic Server, including MVISUALIZE~\cite{Bilbao,PerezMato2015}.

\section{Crystal structure}
\label{str}

The nuclear diffraction patterns are described by the cubic space group $P2_13$ (No.~198), with Mn and Si at $4a$ positions $(x,x,x)$. Figure~\ref{dpatt} compares the merged 2--6~K pattern measured at $\lambda=1.155$~\AA\ in HI mode with the 55~K pattern measured at $\lambda=1.494$~\AA\ in HR mode. With both sites fully occupied, the short-wavelength refinement gives $x_{\rm Mn}=0.86358(9)$ and $x_{\rm Si}=0.15454(7)$, isotropic displacement parameters $B_{\rm Mn}=0.103(9)$ and $B_{\rm Si}=0.094(8)$~\AA$^2$, and $a=4.55109(7)$~\AA.

\begin{figure*}[!t]
  \begin{center}
    \includegraphics[width=0.72\textwidth]{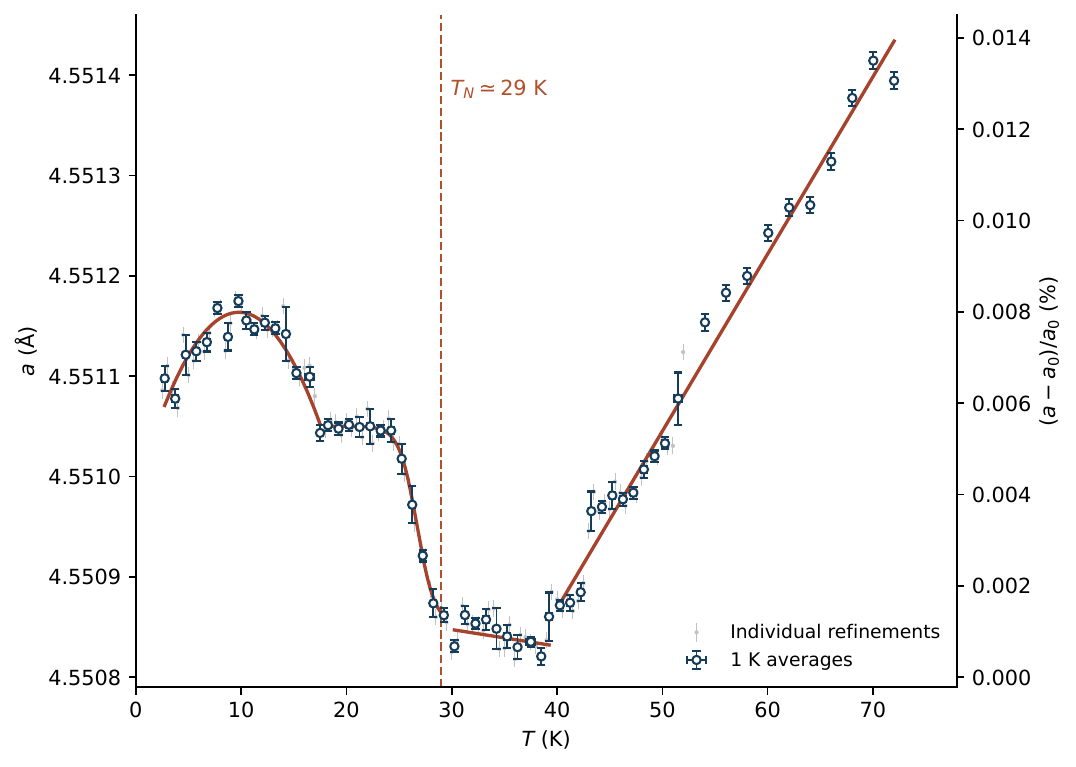}
  \end{center}
\caption{
Temperature dependence of the lattice constant, compiled from individual diffraction patterns at $\lambda=1.155$~\AA. Grey points show individual refinements; open symbols are averages in fixed 1~K bins. Vertical bars include excess scatter within each bin, and horizontal bars combine the within-bin temperature spread with the temperature variations. The solid lines are guides to the eye. The dashed line marks the literature value $T_N\simeq29$~K. The right-hand axis shows the relative change $100(a-a_0)/a_0$ in percent, with $a_0=4.5508$~\AA.
}
\label{c_of_T}
\end{figure*}

The HR patterns at 1.8 and 55~K show no resolved peak splitting. At high angles, the peaks are as narrow as those described by the Na$_2$Al$_2$Ca$_3$F$_{14}$ (NAC) resolution calibration standard~\cite{Gozzo2006NAC,HRPTDataAnalysis}, with no positive strain broadening required. The refined corrections to the Gaussian $\tan^2\theta$ coefficient in the Caglioti width function~\cite{Hewat1975} are $U=-0.00382(35)$ and $-0.00401(34)$~deg$^2$ at 1.8 and 55~K, respectively, both measured in HR mode. These are empirical corrections to the instrumental profile, with $V$ and $W$ also refined, rather than physical negative strains. The mean HR lattice parameter is determined with a formal relative precision better than $10^{-5}$, with $\sigma_a/a\simeq5.5\times10^{-6}$. This uncertainty and the small scatter of the HI temperature-scan values quantify the precision of the mean cubic cell (Fig.~\ref{c_of_T}). 

Selecting a single propagation-vector arm along $\langle111\rangle$ lowers the magnetic symmetry to trigonal and, through magnetoelastic coupling, is generally expected to induce a rhombohedral distortion. No such distortion is detected in our HR data. The retained cubic metric therefore provides a structural motivation for the cubic magnetic superspace model. Distinguishing it conclusively from a single-arm state requires comparing the theoretically predicted magnetostrictive distortion with a quantitative experimental detection limit.

For comparison, our study of isostructural MnGe resolved an orthorhombic structure, $P2_12_12_1$ (No.~19), in the magnetically ordered state below $T_N\simeq170$~K~\cite{PomjakushinMnGe2023}. At 2~K, the relative spread of the lattice parameters is $(c-a)/\bar a\simeq1.6\times10^{-3}$, where $\bar a=(a+b+c)/3$. The resolved distortion favoured orthorhombic magnetic superspace descriptions over the cubic model. No analogous splitting of this size is resolved in MnSi. However, even assuming similar magnetoelastic coupling strengths, the induced strain also depends on the elastic stiffnesses and the magnitude and symmetry of the magnetic order parameter. The MnGe result therefore illustrates the value of structural information for distinguishing magnetic models, but does not provide a quantitative prediction of the rhombohedral distortion in a single-arm MnSi state.

Site occupancies were assessed using the $\lambda=1.155$~\AA\ pattern, which extends to $Q\simeq10.7$~\AA$^{-1}$. This wide $Q$ range helps separate occupancy effects from those of atomic displacement parameters (ADPs): occupancy scales the atomic scattering contribution, whereas ADPs progressively attenuate it with increasing $Q$. With the Si site kept fully occupied, the Mn occupancy refines to $0.965(14)$. Allowing this parameter to vary gives only a small improvement: $\chi^2$ decreases from 3.9281 to 3.9247 and $R_{\rm wp}$ from $4.32\%$ to $4.31\%$. Correlated profile residuals persist, and systematic errors may affect the refined occupancy; the apparent Mn deficit therefore does not establish the presence of vacancies. We describe the sample as near-stoichiometric and retain fully occupied sites as the reference model in Fig.~\ref{dpatt}; the refinements do not prove exact 1:1 stoichiometry. This qualification is relevant for MnSi: studies of controlled starting compositions identify Mn vacancies and antisite disorder, with changes in the magnetic transition, even when bulk composition measurements appear stoichiometric within their uncertainty~\cite{Reiner2016,OuYang2016}.

The lattice constant has a broad maximum near 10~K (Fig.~\ref{c_of_T}), decreases on warming towards the magnetic transition, and has a shallow minimum in the 30--40~K range before increasing again. The contraction on warming towards $T_N$, followed by expansion above the transition region, is qualitatively consistent with the thermal-expansion measurements of Matsunaga et al.~\cite{Matsunaga1982}. Their results do not, however, establish the broad low-temperature maximum observed in our diffraction scan. Dilatometry resolves a small discontinuity near 29~K superposed on a broader thermal-expansion anomaly~\cite{Stishov2007,Petrova2016}. Our diffraction scan is consistent with a nonmonotonic lattice response, but does not resolve this small discontinuity or establish a separate transition near 10~K. Small variations of instrumental wavelength can also affect changes of this size and are not included in the pointwise refinement errors.

The measured $a(T)$ provides a practical guide for selecting reference temperatures for magnetic subtraction. In particular, 54~K is above $T_N$ and has a lattice constant close to that near base temperature, reducing shifts of the nuclear peaks between the two patterns. Matching the lattice constants improves cancellation; it does not guarantee identical intensities or line shapes.

\section{Magnetic structure determination and discussion}
\label{mag_str}
The magnetic satellites in MnSi lie close to much stronger nuclear Bragg reflections. We therefore use the difference patterns between the base-temperature and 54~K diffraction patterns measured on HRPT for the magnetic structure determination. This subtraction suppresses the nuclear contribution, although imperfect cancellation can leave residual features that must be considered when assessing the magnetic fits. 
The subtraction also suppresses scattering from minor impurity phases, provided that their diffraction contribution remains approximately unchanged between base temperature and 54~K.

The magnetic signal requires percent-level discrimination against the nuclear scattering. Across H1--H3, the complete $\{110\}$, $\{111\}$ and $\{210\}$ satellite groups have integrated areas of approximately 1.6\%, 1.1--1.4\% and 1.9--2.1\%, respectively, of the corresponding 54~K nuclear peak areas. The ratio is larger, about 8\%, for the weaker nuclear $\{211\}$ reflection. For a magnetic contribution of 1--2\%, measuring its area to 10\% precision requires residual nuclear contributions to be controlled at roughly 0.1--0.2\% of the nuclear area. This motivates long counting times, matched-temperature references and comparison of independent measurements.

The three HRPT datasets provide independent tests of the magnetic intensity distribution. No additional peak near $51.5^\circ$ or negative-flank features are resolved in H1--H3. Such features, observed in some measurements with both HRPT and DMC, are considered separately in Sec.~\ref{sec:T_dependence} and are not assigned to intrinsic MnSi magnetic satellites.

\subsection{Le Bail fits and weighted profile decomposition}
\label{sec:lebail}

We first tested the indexing of the magnetic difference patterns by le Bail profile fitting in {\tt FULLPROF}~\cite{Fullprof}. In this procedure the integrated reflection intensities are extracted without imposing a magnetic structure-factor model. The satellite positions are specified in the parent cubic reciprocal basis as $\mathbf{H}\pm\mathbf{k}$, where $\mathbf{H}$ is a reciprocal-lattice vector of the parent primitive cubic lattice and $\mathbf{k}$ belongs to the star of $\mathbf{k}_1=[k,k,k]$ along the cubic $\langle111\rangle$ directions (the $\Lambda$ line, labelled LD). We follow the established nomenclature for magnetic irreducible representations (irreps) and magnetic superspace groups (MSSGs)~\cite{Bilbao,isod}.

\begin{figure*}[!t]
\centering
\includegraphics[width=\textwidth]{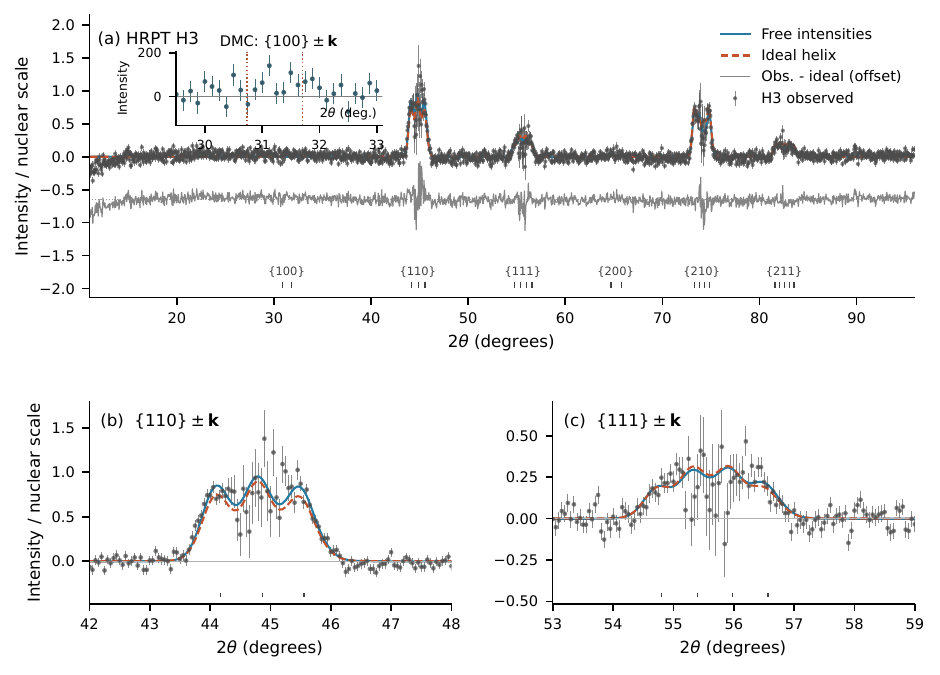}
\caption{HRPT H3 base-minus-54~K profile: (a) full fitted interval; (b,c) enlarged $\{110\}$ and $\{111\}$ satellite groups. Points show measured intensities with counting-error bars; blue and dashed red curves are the weighted free-intensity and ideal-helix fits. The same ideal-fit background is removed from all curves and intensities normalized by the scale factor refined from the 54 K nuclear pattern. The ideal residual is offset in (a). The nominal goodness-of-fit values are $\chi^2=1.3590$ and 1.3887, respectively. The inset shows the DMC difference near the absent $\{100\}\pm\mathbf k$ groups, with its fixed background removed; vertical dotted lines mark their expected positions. The inset retains DMC intensity units and its original counting errors. Short ticks in the main panels mark the indexed satellite positions at the fitted H3 $k$.}
\label{mpatt}
\end{figure*}

The absence of resolved magnetic peaks at the first two satellite groups around $\{100\}$, expected near $2\theta\simeq30.8^\circ$ and $31.7^\circ$, provides a very strong constraint. The DMC data offer a particularly sensitive check in this region [inset of Fig.~\ref{mpatt}(a)]. Neither group is resolved above the local background within experimental sensitivity. Writing $I_{100}$ for the summed area of both first satellite groups, a local weighted fit gives $I_{100}/I_{110}=0.0064(39)$, consistent with zero. Here $I_{110}$ is the complete $\{110\}$ satellite area.

Le Bail extraction updates intensities by iterative partitioning~\cite{Fullprof}; for weak reflections, counting noise and the allocation of intensity between overlapping peaks and background can leave small positive areas without a resolved peak. We therefore use direct weighted whole-pattern decomposition, with satellite intensities and background optimized simultaneously, as the quantitative reference for comparing magnetic models (Sec. S4 A).

FullProf supplied the calculated satellite profiles, with lattice, angular and profile parameters fixed to each dataset's 54~K nuclear refinement and $k$ fixed to its ideal-helix estimate: $0.01433(19)$, $0.01520(20)$ and $0.01467(12)$ for H1, H2 and H3, respectively, in reciprocal-lattice units. After imposing the two $\{100\}$ extinctions, 296 reflection entries form 18 groups of coincident reflections: within each group, the reflections have the same calculated $d$-spacing and hence the same $2\theta$ for the cubic lattice and chosen propagation-vector star. One fitted coefficient describes their combined intensity. Writing $P_g(2\theta)$ for the fixed calculated profile of group $g$, the fitted intensity is $ I_i^{\rm calc}=\sum_{g=1}^{18}c_g P_g(2\theta_i)+b(2\theta_i). $ Here $i$ runs over all measured angular bins, whereas $g$ labels the groups. The function $b(2\theta)$ is a single quadratic background across the entire fitted range, with the same functional form as in the ideal-helix FullProf fit. The 18 group coefficients and three background coefficients were optimized simultaneously by weighted least squares. The measured points around the two extinct $\{100\}$ groups remained included in the fit. Figure~\ref{mpatt} compares this decomposition with the ideal helix for H3, the dataset with the smallest statistical uncertainties.

The free-intensity decomposition provides a reference for the best fit quality attainable under the adopted calibration, profile, background and extinction constraints, without imposing additional magnetic intensity relations between the retained groups. 

The DMC le Bail refinement gives the propagation-vector component $k=0.01510(10)$.

\subsection{Symmetry analysis and magnetic models}
\label{symmetry}

The parent space group $P2_1 3$ (No.~198) has three magnetic irreducible representations (irreps) associated with the LD line $(k,k,k)$. The pure mLD1 model fits the measured magnetic profiles poorly and has an independent difficulty: threefold symmetry forces its structure factor at $\mathbf Q=\mathbf k_1$ to be longitudinal. Its perpendicular component therefore vanishes, in conflict with the observed small-angle satellites~\cite{Ishikawa1976,Grigoriev2006}. We consequently consider mLD2 and its complex-conjugate partner mLD3, corresponding to opposite helicities that cannot be distinguished by these unpolarized powder measurements.

The propagation-vector star consists of four symmetry-equivalent wave vectors $\mathbf{k}_1=(k,k,k)$, $\mathbf{k}_2=(k,-k,-k)$, $\mathbf{k}_3=(-k,k,-k)$ and $\mathbf{k}_4=(-k,-k,k)$, which satisfy the linear relation $\sum_i\mathbf{k}_i=0$. Consequently, only three propagation vectors are linearly independent, although all four arms are required to preserve the full cubic symmetry of the propagation-vector star. For the selected mLD2 order parameter, however, the conventional $3+3$D MSSG generated by ISODISTORT at a fixed phase sum is orthorhombic, as explained below.

For a single active arm $\mathbf k_1$, the mLD2 model has rhombohedral MSSG $R3.1'(0,0,g)ts$ (No.~146.1.22.2.m11.2). In its standard $R$-centred hexagonal setting, the basis is
\begin{equation}
\begin{gathered}
\mathbf a_h=\mathbf b_c-\mathbf c_c,\qquad
\mathbf b_h=\mathbf a_c-\mathbf b_c,\\
\mathbf c_h=-\mathbf a_c-\mathbf b_c-\mathbf c_c,\\
\mathbf k_h=(0,0,g),\qquad g=-3k,
\end{gathered}
\label{eq:R3_standard_basis}
\end{equation}
with zero origin shift. The subscripts $c,h$ denote parent cubic and hexagonal axes; $a_h=\sqrt2a_c$, $c_h=\sqrt3a_c$ and $\gamma_h=120^\circ$. The Mn $4a$ orbit splits into $3a$ and $9b$, allowing two and six independent real first-harmonic coefficients, respectively; fixing the unobservable common phase leaves seven magnetic variables. 

The mLD1 exclusion, rhombohedral conversion conventions and matched $R3$/cubic refinements are detailed in Sec.~S5.

Following a suggestion by B.~J.~Campbell~\cite{Campbell_priv}, the order-parameter direction $(a,b;a,b;a,b;a,b)$ for irrep mLD2LE2 remains invariant under the threefold rotation about the cubic $[111]$ axis only when the rotational part is combined with the superspace phase translation
$(-1/3,-1/3,-1/3,-1/3)$.
Importantly, these superspace phase translations are associated with the four \emph{active} propagation vectors rather than with the three independent modulation coordinates.

Using this observation, we constructed the cubic $3+4$D superspace operators from the generators of the parent space group and the representation matrices of the 8-dimensional mLD2LE2 irrep. Each matrix was applied to the order-parameter direction $(a,b;a,b;a,b;a,b)$, and compensating translations of the four internal coordinates were introduced where required to restore invariance. For the threefold rotation about $[111]$, the phase factor $\exp(2\pi i/3)$ is compensated by the internal translation $(2/3,2/3,2/3,2/3)$, equivalent modulo one to the translation given above. This procedure yields the operators in Table~\ref{tab:ssg_ops}.

Operations 2 and 3 generate the 12 listed representatives modulo lattice translations. Their rotational parts, denoted by $s$ and $t$, satisfy $s^2=t^3=(st)^3=e$, the standard presentation of the tetrahedral group $A_4$. The superspace operators were verified to close under composition modulo lattice translations. The complete magnetic group also includes the time-reversal operation combined with a half-period shift of all four internal coordinates, specified in the table caption. We list the operators explicitly because this cubic $3+4$D construction is not tabulated in the ISOTROPY~\cite{isot} database.

\begin{widetext}
\noindent\begin{minipage}{\textwidth}
\captionof{table}{Symmetry operations of the cubic $3+4$-dimensional magnetic superspace group for the irrep mLD2LE2 with order parameter $(a,b;a,b;a,b;a,b)$. The first three coordinates correspond to the crystallographic coordinates, whereas the remaining four superspace coordinates correspond to the four active propagation vectors of the star. The last entry denotes the time-reversal character of the operation. The complete magnetic superspace group additionally contains the anti-unitary centering operation
$(x_1,x_2,x_3,
x_4+\tfrac12,
x_5+\tfrac12,
x_6+\tfrac12,
x_7+\tfrac12),-1.
$
}
\label{tab:ssg_ops}
\centering
\small
\begin{tabular}{cl}
\hline
No. & Superspace operation \\
\hline
1 &
$(x_1,x_2,x_3,x_4,x_5,x_6,x_7,+1)$ \\

2 &
$(-x_1+\frac12,-x_2,x_3+\frac12,x_7,x_6,x_5,x_4,+1)$ \\

3 &
$(x_3,x_1,x_2,
x_4+\frac23,
x_7+\frac23,
x_5+\frac23,
x_6+\frac23,+1)$ \\

4 &
$(x_3+\frac12,-x_1+\frac12,-x_2,
x_7+\frac23,
x_4+\frac23,
x_6+\frac23,
x_5+\frac23,+1)$ \\

5 &
$(-x_3+\frac12,-x_1,x_2+\frac12,
x_6+\frac23,
x_5+\frac23,
x_7+\frac23,
x_4+\frac23,+1)$ \\

6 &
$(x_2,x_3,x_1,
x_4+\frac13,
x_6+\frac13,
x_7+\frac13,
x_5+\frac13,+1)$ \\

7 &
$(-x_3,x_1+\frac12,-x_2+\frac12,
x_5+\frac23,
x_6+\frac23,
x_4+\frac23,
x_7+\frac23,+1)$ \\

8 &
$(-x_2,x_3+\frac12,-x_1+\frac12,
x_7+\frac13,
x_5+\frac13,
x_4+\frac13,
x_6+\frac13,+1)$ \\

9 &
$(x_2+\frac12,-x_3+\frac12,-x_1,
x_6+\frac13,
x_4+\frac13,
x_5+\frac13,
x_7+\frac13,+1)$ \\

10 &
$(-x_2+\frac12,-x_3,x_1+\frac12,
x_5+\frac13,
x_7+\frac13,
x_6+\frac13,
x_4+\frac13,+1)$ \\

11 &
$(-x_1,x_2+\frac12,-x_3+\frac12,
x_6,x_7,x_4,x_5,+1)$ \\

12 &
$(x_1+\frac12,-x_2+\frac12,-x_3,
x_5,x_4,x_7,x_6,+1)$ \\

\hline
\end{tabular}
\end{minipage}
\end{widetext}

For the cubic mLD2 four-arm construction we introduce the extended, non-tabulated symbol $P2_1 3\,1'(a,a,a)\,0{-t}s\allowbreak (a,-a,-a)\,0{-t}s\allowbreak (-a,a,-a)\,0{-t}s\allowbreak (-a,-a,a)\,0{-t}s$. Here the translation codes are standard $t=1/3$ and $s=1/2$. For each arm, the entries $0,-t,s$ specify the internal translations associated with the $2_1$, $3$ and $1'$ generators, respectively. The symbol denotes the extended four-modulation family specified by Table~\ref{tab:ssg_ops}; its relation to the conventional fixed-phase-sum description is discussed below.

For the selected order parameter, the conventional $3+3$D description generated by ISODISTORT uses three independent modulation coordinates for four active arms. Since $\sum_i\mathbf k_i=0$, the phase sum $\Sigma=x_4+x_5+x_6+x_7$ is constant in space on a chosen three-dimensional section of superspace representing the physical structure. The internal coordinates are expressed in cycles, so the dependent coordinate can be written as
$
x_7=\Sigma-x_4-x_5-x_6\pmod{1}.
$
Eliminating $x_7$ therefore retains the fourth helix with a dependent phase; it does not remove its modulation. A spatial translation cannot change $\Sigma$, although different four-arm configurations may have different values of this constant.

An operation must preserve $\Sigma$ modulo an integer to act within this fixed phase-sum description. For example, the threefold operation No.~3 of Table~\ref{tab:ssg_ops} adds $2/3$ to each of the four internal coordinates, changing their sum by $8/3\equiv2/3$ modulo one; it is therefore excluded. Retaining the operations that preserve the sum reproduces the orthorhombic MSSG No.~19.3.146.7.m26.2 obtained with the ISODISTORT model, with basic space group $P2_1 2_1 2_1$. The extended $3+4$D construction retains the phase transformation of every active arm and the cubic superspace symmetry of the family of phase sections; some operations map between sections with different phase sums. It is therefore a useful common description of the four-arm configurations considered here.

For the Mn atom occupying the $4a$ Wyckoff position $(x,x,x)$, the complete crystallographic orbit is generated by the following four superspace operations:
\begin{widetext}
\begin{equation}
\begin{aligned}
Mn_1 &: (x_1,x_2,x_3,x_4,x_5,x_6,x_7),\\
Mn_2 &: (x_1+\tfrac12,-x_2+\tfrac12,-x_3,
        x_5,x_4,x_7,x_6),\\
Mn_3 &: (-x_1,x_2+\tfrac12,-x_3+\tfrac12,
        x_6,x_7,x_4,x_5),\\
Mn_4 &: (-x_1+\tfrac12,-x_2,x_3+\tfrac12,
        x_7,x_6,x_5,x_4).
\end{aligned}
\label{eq:Mnorbit}
\end{equation}
\end{widetext}

The four superspace operations in Eq.~\ref{eq:Mnorbit} relate the representative Mn site to all four symmetry-related Mn sites, specifying both their spatial coordinates and the corresponding transformations of the four internal coordinates. To derive the magnetic modulation from Table~\ref{tab:ssg_ops}, we must also impose the constraints from operations that leave a representative site invariant. At $(x,x,x)$, the threefold rotation leaves the Mn position and the propagation vector $\mathbf k_1$ unchanged, while cyclically permuting the Cartesian moment components and the remaining three propagation-vector arms. Its accompanying internal translations impose phase relations between the modulation functions. These constraints determine the allowed modulation at the representative site; the modulations at the other three sites then follow from the orbit-generating operations of Eq.~\ref{eq:Mnorbit}.

Thus, the magnetic superspace symmetry determines the allowed magnetic modulation on the four Mn sites. Using the notation introduced in Eq.~(\ref{eq:Mnorbit}), the magnetic moments can be written in terms of four real modulation amplitudes $m_i$ and four phases $\alpha_i$ $(i=0,\ldots,3)$, and the fixed phase shifts $\pm1/3$ originating from the action of the threefold symmetry operator. For compactness, we introduce $\varphi_{ji}=x_j+\alpha_i$, where $j=4,\ldots,7$ labels the superspace coordinates associated with the propagation vectors $\mathbf{k}_1,\ldots,\mathbf{k}_4$, and $i=0,\ldots,3$ labels the modulation amplitude and its associated phase. The indices $j$ and $i$ label different objects: an arm coordinate and an amplitude--phase pair, respectively. All phases are expressed in cycles, and $\cos u$ abbreviates $\cos(2\pi u)$.

\begin{widetext}
\begin{subequations}
\label{eq:general_modulation}
\begin{align}
\mathbf{M}_{\mathrm{Mn1}} &=
\begin{pmatrix}
 m_0\cos\varphi_{40}
+m_1\cos\varphi_{51}
+m_3\cos(\varphi_{63}+\frac13)
+m_2\cos(\varphi_{72}-\frac13)
\\
 m_0\cos(\varphi_{40}+\frac13)
+m_2\cos\varphi_{52}
+m_1\cos(\varphi_{61}+\frac13)
+m_3\cos(\varphi_{73}-\frac13)
\\
 m_0\cos(\varphi_{40}-\frac13)
+m_3\cos\varphi_{53}
+m_2\cos(\varphi_{62}+\frac13)
+m_1\cos(\varphi_{71}-\frac13)
\end{pmatrix},
\label{eq:Mn1}
\\[2ex]
\mathbf{M}_{\mathrm{Mn2}} &=
\begin{pmatrix}
 m_1\cos\varphi_{41}
+m_0\cos\varphi_{50}
+m_2\cos(\varphi_{62}-\frac13)
+m_3\cos(\varphi_{73}+\frac13)
\\
 -m_2\cos\varphi_{42}
-m_0\cos(\varphi_{50}+\frac13)
-m_3\cos(\varphi_{63}-\frac13)
-m_1\cos(\varphi_{71}+\frac13)
\\
 -m_3\cos\varphi_{43}
-m_0\cos(\varphi_{50}-\frac13)
-m_1\cos(\varphi_{61}-\frac13)
-m_2\cos(\varphi_{72}+\frac13)
\end{pmatrix},
\label{eq:Mn2}
\\[2ex]
\mathbf{M}_{\mathrm{Mn3}} &=
\begin{pmatrix}
 -m_3\cos(\varphi_{43}+\frac13)
-m_2\cos(\varphi_{52}-\frac13)
-m_0\cos\varphi_{60}
-m_1\cos\varphi_{71}
\\
 m_1\cos(\varphi_{41}+\frac13)
+m_3\cos(\varphi_{53}-\frac13)
+m_0\cos(\varphi_{60}+\frac13)
+m_2\cos\varphi_{72}
\\
 -m_2\cos(\varphi_{42}+\frac13)
-m_1\cos(\varphi_{51}-\frac13)
-m_0\cos(\varphi_{60}-\frac13)
-m_3\cos\varphi_{73}
\end{pmatrix},
\label{eq:Mn3}
\\[2ex]
\mathbf{M}_{\mathrm{Mn4}} &=
\begin{pmatrix}
 -m_2\cos(\varphi_{42}-\frac13)
-m_3\cos(\varphi_{53}+\frac13)
-m_1\cos\varphi_{61}
-m_0\cos\varphi_{70}
\\
 -m_3\cos(\varphi_{43}-\frac13)
-m_1\cos(\varphi_{51}+\frac13)
-m_2\cos\varphi_{62}
-m_0\cos(\varphi_{70}+\frac13)
\\
 m_1\cos(\varphi_{41}-\frac13)
+m_2\cos(\varphi_{52}+\frac13)
+m_3\cos\varphi_{63}
+m_0\cos(\varphi_{70}-\frac13)
\end{pmatrix}.
\label{eq:Mn4}
\end{align}
\end{subequations}
\end{widetext}

Equations~\eqref{eq:Mn1}--\eqref{eq:Mn4} describe the most general first-harmonic modulation allowed by the cubic $3+4$D superspace symmetry. Additional relations between the amplitudes $m_i$ and phases $\alpha_i$ select the models considered below. A model with fewer active arms is obtained by retaining only the terms containing the corresponding internal coordinates on all four Mn sites; this is distinct from setting one of the $m_i$ to zero.

For a single active arm, the resulting modulation is equivalent to the standard rhombohedral description $R3.1'(0,0,g)ts$ given above. Retaining the parent operations that preserve arms 1 and 2 gives a monoclinic $P2_1$ MSSG, whereas three active arms give a rhombohedral $R3$ MSSG. The choices $\{1,2,3\}$ and $\{2,3,4\}$ are symmetry-related three-arm domains. Explicit transformations establish agreement with the tabulated groups, including the internal-coordinate operations and centering translations.

MSSG symmetry alone permits 24 real first-harmonic Mn coefficients for each of the two-arm $P2_1$, three-arm $R3$, and four-arm $3+3$D $P2_12_12_1$ descriptions. Our models retain the eight real parameters of Eq.~\ref{eq:general_modulation}: deleting arms or eliminating the dependent coordinate $x_7$ preserves its component and intersite phase relations. These inherited relations impose additional parameter constraints beyond those of the respective MSSG; they do not imply additional symmetries of the reduced model. The single-arm $R3$ model instead already has eight coefficients allowed by its MSSG. The ideal-helix relations impose further restrictions within these eight-parameter families. The transformations, explicit coefficient relations and counting, and ideal-field expressions and illustrations of the $3+2$D, $3+3$D and $3+4$D constructions are collected in Sec.~S9.

The apparent special role of $m_0$ in the representative-site modulation reflects the different action of the local threefold symmetry on the four-arm star. In contrast to the three-arm star of MnGe~\cite{PomjakushinMnGe2023}, which is cyclically permuted as a whole by the local threefold rotation, the four-arm star of MnSi decomposes into one invariant arm, parallel to the local $[111]$ axis, and a three-arm cycle. Consequently, a modulation written for a fixed Mn site cannot be manifestly symmetric with respect to all four amplitudes. The cubic superspace symmetry makes the four Mn sites equivalent: any one can be chosen as the representative site, with a corresponding relabelling of the propagation-vector arms.

The special role of the modulation associated with the arm parallel to the local threefold axis is not unique to mLD2. In the cubic $3+3$D mLD1 MSSG generated by ISODISTORT, the first arm is likewise invariant under the local threefold operation, while the remaining three arms form a cycle. The essential difference is that mLD1 requires no additional internal phase translation and is therefore compatible with a tabulated cubic $3+3$D MSSG based on three independent modulation coordinates.

\subsection{Ideal helix model}
\label{magstr}

\begin{figure*}[!t]
\centering
\includegraphics[width=\textwidth]{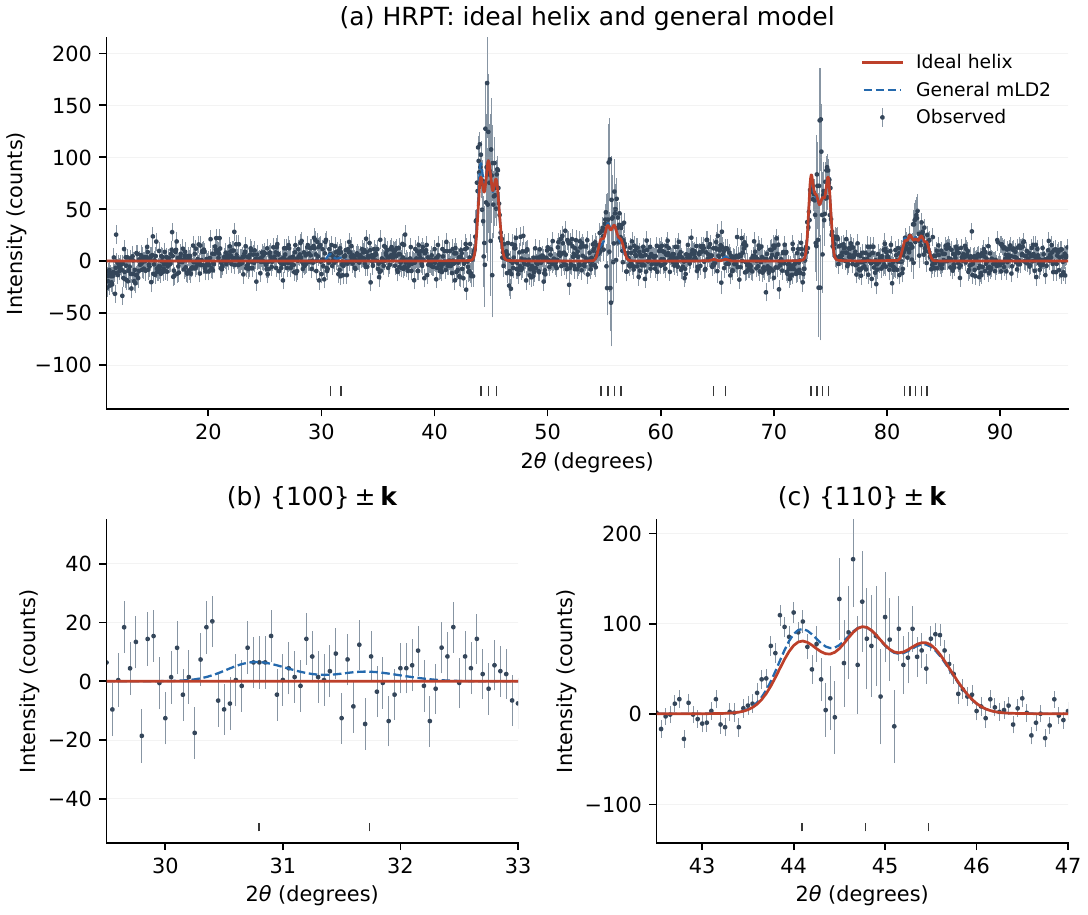}
\caption{Test of the ideal-helix constraint using the HRPT H2 dataset (July 2025). The one-parameter ideal helix (red) and the general seven-parameter mLD2 refinement (blue dashed) use identical profile settings and reflection support. Points with error bars show the base-temperature minus 54~K intensities, with the same fixed background subtracted from the measured intensities and calculated profiles. Panel (a) shows the fitted range; panels (b) and (c) enlarge the first $\{100\}$ satellite groups and the strongest $\{110\}$ group. The ideal helix extinguishes both first satellites. Short vertical ticks mark satellite positions, including the two $\{100\}$ groups allowed before imposing the ideal-helix constraint. The fitted profiles give $\chi^2=1.1785$ and 1.1693, respectively.}
\label{fig:ideal_helix_HRPT}
\end{figure*}

The absence of both satellite groups around $\{100\}$ provides the central constraint on the magnetic structure. Contributions from different arms add as intensities, so an exact extinction requires each populated contribution to vanish separately. Within the model of Sec.~\ref{symmetry}, simultaneous extinction of both groups selects an ideal transverse circular helix for every active arm. In the notation of Eq.~\ref{eq:general_modulation}, the corresponding constraints are
\begin{equation}\begin{gathered}
 m_0=m_1=m_2=m_3=m,\\
 (\alpha_1-\alpha_0,\alpha_2-\alpha_0,\alpha_3-\alpha_0)\\
 {}=(0,-1/6,+1/6)\pmod1.
\end{gathered}\label{eq:ideal_helix_polar}
\end{equation}
The derivation and phase conventions are given in Sec.~S6. The measurements are consistent with the exact extinction predicted by the ideal helix, but cannot exclude small departures that produce intensities below our detection limit. The exact-extinction limit defined by Eq.~\ref{eq:ideal_helix_polar} is tested against the full magnetic diffraction pattern.

With these constraints and the fixed symmetry phases of Eq.~\ref{eq:general_modulation}, all four Mn sites share the same transverse circular modulation function for a given arm. Their contributions cancel at $(100)\pm\mathbf k$, giving $\mathbf F=0$ independently of the propagation-vector component $k$ and Mn coordinate $x$. The common phase translates a single helix and is unobservable in the satellite intensities. For the four-arm superposition, it can change the real-space texture, as discussed in Sec.~\ref{sec:continuum_topology}. Fixing this diffraction-invisible phase leaves only $m$ to refine, with the propagation vector and nuclear scale held fixed.

\begin{figure*}[!t]
\centering
\includegraphics[width=\textwidth]{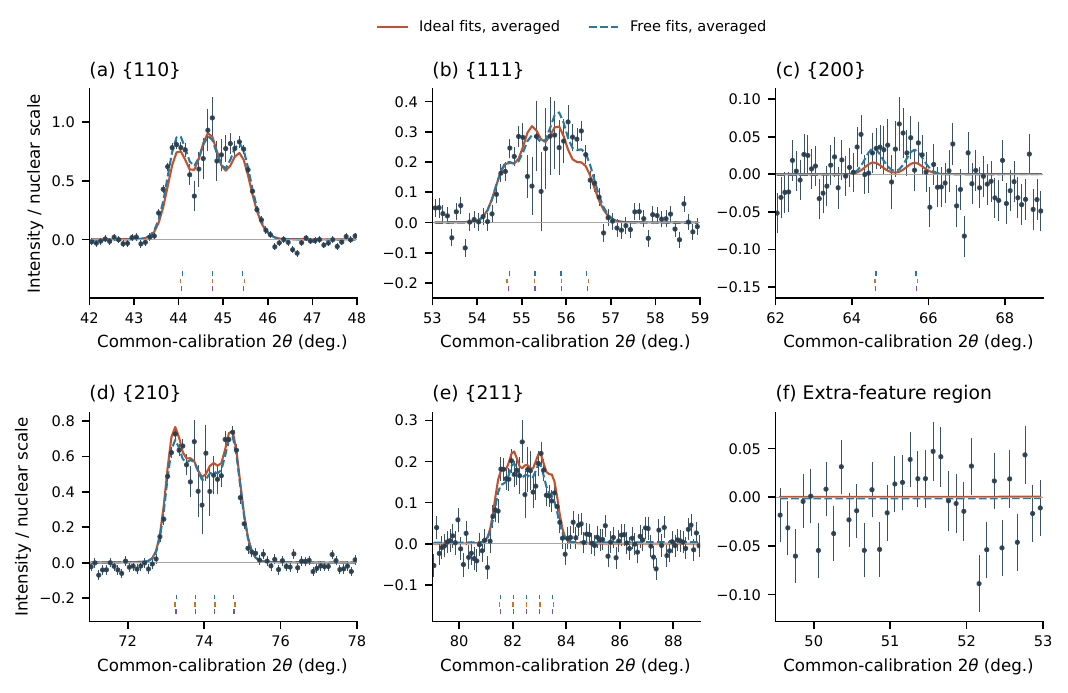}
\caption{Display-only weighted average of H1, H2 and H3 in $0.10^\circ$ common-calibration bins. The three campaigns contribute approximately 360~h of counting in total, including about 202~h at base temperature and 157~h at 54~K. Red and dashed blue lines are the independently fitted ideal and free-intensity profiles averaged with the same weights. Different instrumental profiles and propagation components remain present in the average; no refinement is performed on this histogram. Tick rows in (a)--(e), from top to bottom, mark the satellite positions for H1 (blue), H2 (orange) and H3 (purple), transformed to the same angular calibration. No MnSi satellite is indexed in (f).}
\label{fig:HRPT_display_average}
\end{figure*}

The ideal model describes the principal magnetic features in all three independent HRPT difference patterns (Fig.~\ref{fig:HRPT_display_average}). Table~\ref{tab:ideal_all} gives the single-arm amplitude $m_{1k}$ and the powder-equivalent four-arm amplitude $m_{4k}=m_{1k}/2$ per arm. This factor of two compensates for adding the intensities of four symmetry-related arms. Here $m$ is the Cartesian modulation amplitude in Eq.~\ref{eq:general_modulation}; the constant moment magnitude of a single ideal helix is $\sqrt{3/2}\,m$.

The extinction condition thus applies both to the single-arm $3+1$D structure and to each arm of the cubic $3+4$D model, without selecting the number of simultaneously active arms. A coherent superposition of four ideal helices need not have a constant total moment magnitude, although its powder profile is identical to that of the appropriately scaled single-arm model.

\begin{figure*}[!t]
\centering
\includegraphics[width=\textwidth]{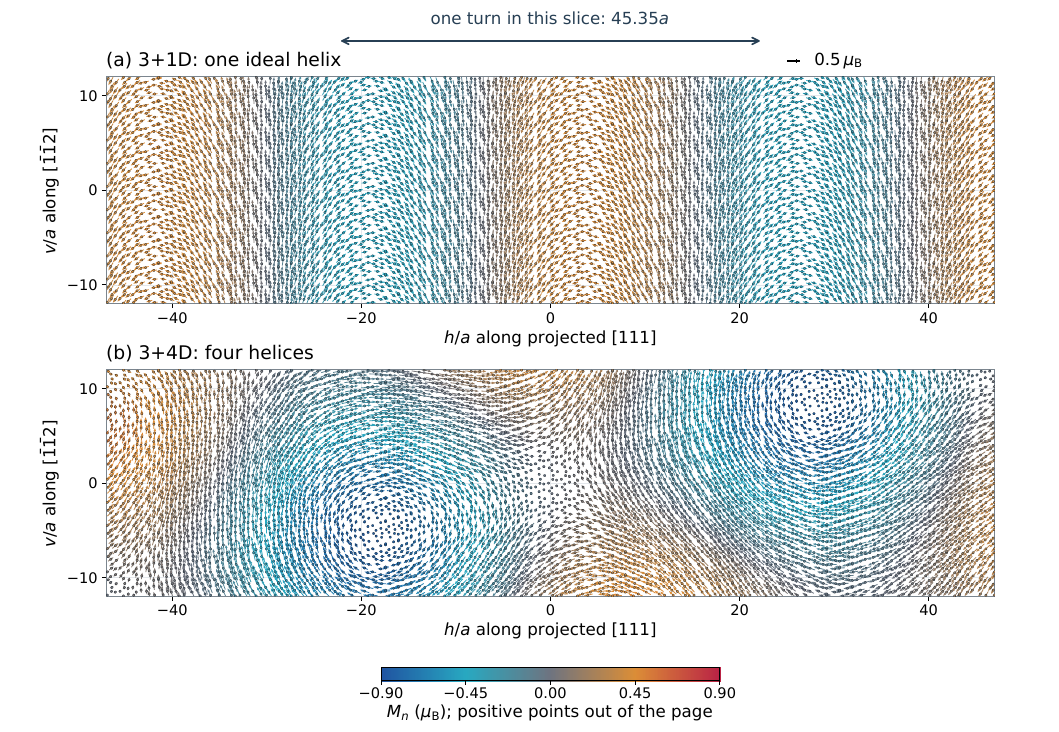}
\caption{Real-atom views of the ideal (a) single-arm $3+1$D and (b) four-arm $3+4$D structures from Eq.~\ref{eq:general_modulation}. Each panel shows the same Mn sites in a $94a\times24a$ slice of thickness $a/\sqrt2$. The slice is viewed along its normal $\hat{\mathbf n}=\tfrac12\hat{\mathbf k}_1+ \tfrac{\sqrt3}{2}[1\bar10]/\sqrt2$, where $\hat{\mathbf k}_1=[111]/\sqrt3$; thus the viewing direction makes $60^\circ$ with $\mathbf k_1$. The horizontal direction $\hat{\mathbf h}$ is the normalized projection of $\mathbf k_1$ into the slice plane, and the vertical direction is $\hat{\mathbf v}=[\bar1\bar12]/\sqrt6$. These directions satisfy $\hat{\mathbf h}\times\hat{\mathbf v}=\hat{\mathbf n}$. The horizontal axis labelled ``projected $[111]$'' therefore makes $30^\circ$ with the actual propagation direction. Arrows show the in-plane moments; arrow and atom colours encode $M_n=\mathbf M\cdot\hat{\mathbf n}$, positive out of the page. }
\label{fig:ideal_structure}
\end{figure*}

\begin{figure}[tbp]
\centering
\includegraphics[width=\columnwidth]{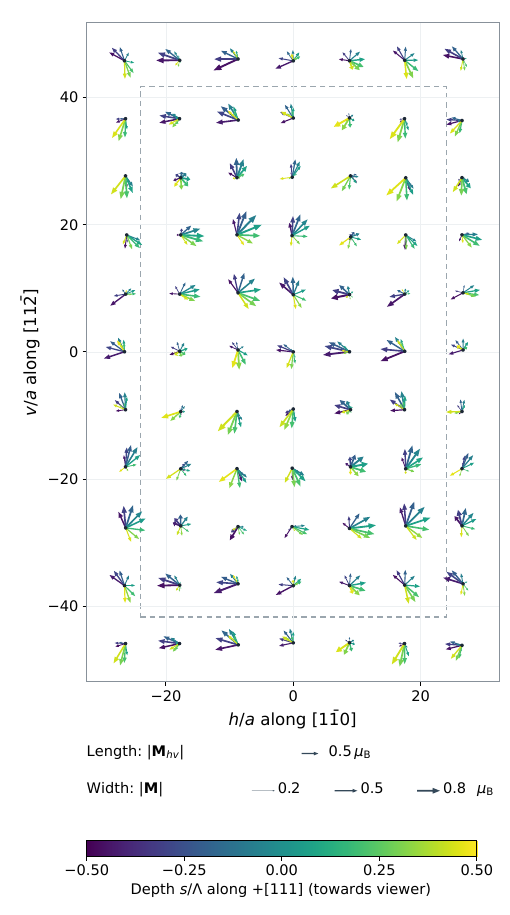}
\caption{Four-arm $3+4$D configuration viewed along $[111]$.
Arrows show 616 actual Mn moments at eight depths along 77 sampled
columns in a slab of thickness $\Lambda=a/(\sqrt3 k)=39.28a$.
Length denotes the projected moment, shaft width its full magnitude,
and colour the depth, positive towards the viewer. Black dots mark
projected columns.
The dashed rectangle is a $48.10a\times83.32a$ transverse modulation
repeat. The moment scale, $k=0.0147$, and illustrative common phase
$\alpha_0=0$ match Fig.~\ref{fig:ideal_structure}.}
\label{fig:ideal_4k_endon}
\end{figure}

Figure~\ref{fig:ideal_structure} compares the moments on the actual Mn sites in an oblique slice spanning just over two helix periods. The view shows both the transverse rotation and its progression along the projected $[111]$ direction. For $1k$, the full moment rotates in a plane perpendicular to the propagation vector with constant length. The $4k$ example superposes four such helices, giving a spatially varying direction and magnitude. 
A complementary projection along $[111]$ is shown in Fig.~\ref{fig:ideal_4k_endon}, with selected atomic depths distinguished by colour and the full moment magnitude encoded by arrow width.

The illustrated common-phase choice $\alpha_0=0$ specifies one real-space configuration and is not determined by the satellite intensities. For this phase choice, the four-arm moment ranges from zero to $4m_{4k}=2m_{1k}=0.874(5)\,\mu_{\rm B}$ per Mn using the averaged HRPT amplitude, while its root-mean-square magnitude is $\sqrt{\langle|\mathbf M|^2\rangle}=\sqrt6\,m_{4k} =0.535(3)\,\mu_{\rm B}$, equal to the constant single-helix moment.

\begin{table*}[tbp]
\caption{Comparison of ideal-helix and free-intensity fits. Amplitudes are Cartesian components in $\mu_{\rm B}$ on each dataset's fixed nuclear scale; the $4k$ value is per arm. Both ideal descriptions give the same powder profile and $\chi^2$. H1 and H2 use $11^\circ<2\theta<96^\circ$ ($N=1699$); H3 uses $20^\circ<2\theta<96^\circ$ ($N=1519$). The HRPT ideal fits vary amplitude, propagation component and three background coefficients ($n_{\rm par}=5$); the weighted free-intensity values use the nominal count $n_{\rm par}=22$. DMC uses $20^\circ$--$88^\circ$, excluding $50^\circ$--$53^\circ$, with only amplitude varied in the ideal fit: its listed $k$ and uncertainty are from the separate Le Bail refinement. The DMC free-intensity entry is the native FullProf Le Bail value ($S/N$), with the $\{100\}$ intensities fixed to zero and their observed bins retained. The DMC row is a complementary comparison, not a fit of comparable quality to HRPT.}
\label{tab:ideal_all}
\setlength{\tabcolsep}{5pt}
\begin{ruledtabular}
\begin{tabular}{lccccc}
Dataset & $k$ (r.l.u.) & $m_{1k}$ & $m_{4k}$ & $\chi^2_{\rm ideal}$ & $\chi^2_{\rm free}$ \\
H1  & $0.01433(19)$ & $0.4410(56)$ & $0.2205(28)$ & 1.0685 & 1.0547 \\
H2  & $0.01520(20)$ & $0.4426(57)$ & $0.2213(29)$ & 1.1343 & 1.1017 \\
H3  & $0.01470(12)$ & $0.4372(33)$ & $0.2186(17)$ & 1.0683 & 1.0258 \\
DMC & $0.01510(10)$ & $0.4729(56)$ & $0.2364(28)$ & 15.9856 & 8.2918 \\
\end{tabular}
\end{ruledtabular}
\end{table*}

The three HRPT amplitudes agree within their uncertainties. An inverse-variance average of the three native FullProf results gives $\overline{m}_{1k}=0.4371(27)\,\mu_{\rm B}$ and supplies the moment uncertainties.

In the standard $R$-centred hexagonal setting of Eq.~\ref{eq:R3_standard_basis}, choosing $\alpha_0=0$ and writing $\mathbf M(t)=\mathbf C\cos(2\pi t)+\mathbf S\sin(2\pi t)$ gives $\mathbf C=(0.3091(19),\,0.6182(38),\,0)\,\mu_{\rm B}$ and $\mathbf S=(-0.5353(33),\,0,\,0)\,\mu_{\rm B}$ for the representatives of both Mn orbits, $3a$ and $9b$. These components are defined along unit crystallographic axes and correspond to the constant moment $0.535(3)\,\mu_{\rm B}$ per Mn. All coefficients are constrained by the single ideal-helix amplitude.

Figure~\ref{fig:HRPT_display_average} shows a weighted display average of all three datasets on a common angular calibration, illustrating their combined counting statistics. The separate data points and ideal profiles are compared in Fig.~S4. Because the calibrations and resolution functions differ, this average is not treated as a single diffraction dataset for refinement. Combined consistency tests fit the three original patterns simultaneously, retaining their separate calibrations, resolution functions and backgrounds, and minimizing the sum of their weighted squared residuals.

For all three HRPT datasets, the ideal-helix $\chi^2$ values approach those of the free-intensity decomposition (Table~\ref{tab:ideal_all}), although their magnetic intensities are described by a single adjustable amplitude. In DMC, substantial residuals remain even after independent optimization of the satellite intensities, so its larger ideal-model $\chi^2$ does not by itself identify a departure from the magnetic constraints.

\subsection{Tests of departures from the ideal helix}
\label{sec:SA}

Simulated annealing (SA)~\cite{kirkpatrick83}, as implemented in FullProf~\cite{Fullprof}, was used to explore the general  mLD2 model of Eq.~\ref{eq:general_modulation}. The searches varied the real and imaginary parts of its four complex Fourier coefficients, retaining the symmetry relations between Mn sites and active arms but without imposing the ideal-helix amplitude and phase constraints. Starting from random coefficient sets, SA permits moves that increase the residual, helping the search escape local minima as the annealing control parameter is progressively reduced. The resulting candidate solutions were subsequently refined by least squares. Fixing the common phase, which is unobservable in these intensities, leaves seven real magnetic parameters.

For the matched H2 test, we also used nonlinear trust-region least
squares from 32 random coefficient sets, evaluating the intensities
through quadratic representations of fixed FullProf profiles. The resulting minima
were refined in FullProf; direct FullProf searches from the same
starting sets also recovered both minima (Sec.~S7).

In the earlier H2 comparison with identical fixed profile, background, propagation vector and reflection support, the ideal model gives $\chi^2=1.1785$ and the general model gives 1.1693 (Fig.~\ref{fig:ideal_helix_HRPT}). The profiles differ little on the scale of the measured intensities. Both the SA solution and the matched least-squares refinement retain relative phases close to the ideal-helix values, but give unequal modulation amplitudes. They therefore suggest possible departures from the ideal amplitude relations. Several coefficient combinations are strongly correlated, and nearby minima remain. The small profile improvement therefore does not provide a robust, uniquely determined distortion of the ideal helix.

The small relative phase shift between Mn orbits proposed by symmetry-constrained muon-spin-rotation analyses~\cite{DalmasDeReotier2016,DalmasDeReotier2018} is a complementary constraint. Agreement with an ideal powder profile and a finite upper limit on the $\{100\}$ satellites do not by themselves exclude such a small deviation. A quantitative comparison requires calculating its satellite intensities in the same phase convention.

In analogous DMC exploratory refinements, substantial profile residuals persist even without magnetic intensity constraints, preventing a unique magnetic interpretation of the improvement in a general model. We therefore use the ideal helix as the reference description. Representative HRPT coefficients and the exploratory DMC comparison are given in Sec.~S7.

\section{Continuous limit and topology}
\label{sec:continuum_topology}

Superspace symmetry alone does not guarantee a single slowly varying first-harmonic field common to all four Mn sublattices. Within Eq.~\ref{eq:general_modulation}, this requires the special amplitude and phase relations of the ideal-helix limit, Eq.~\ref{eq:ideal_helix_polar}. They compensate the site-dependent rotations and phase offsets, as in the MnGe construction~\cite{PomjakushinMnGe2023}, and reduce all four Mn modulation functions to the same function of the internal coordinates. Define, for $j=4,\ldots,7$,
\[
\begin{aligned}
A_j&=\cos\!\left[2\pi(x_j+\alpha_0)\right],\\
A_{\pm j}&=\cos\!\left[2\pi\left(x_j+\alpha_0\pm\tfrac13\right)\right].
\end{aligned}
\]
The common modulation function is
\begin{equation}
\mathbf M=m
\begin{pmatrix}
A_4+A_5-A_6-A_7\\
A_{+4}-A_{+5}+A_{+6}-A_{+7}\\
A_{-4}-A_{-5}-A_{-6}+A_{-7}
\end{pmatrix}.
\label{eq:common_modulation}
\end{equation}
Here $m$ is the common amplitude of Eq.~\ref{eq:ideal_helix_polar}, and $j$ labels the propagation-vector arms. The equality of the four modulation functions holds for arbitrary independent $x_4,\ldots,x_7$; the different Mn positions enter only through the internal-coordinate values at which this common function is evaluated. For any subset of active arms within this ideal-helix family, only the terms belonging to those arms are retained in each component. A common continuous field is obtained by taking a three-dimensional section of this function, as in the MnGe construction~\cite{PomjakushinMnGe2023}.

For the four-arm 3+4D configuration considered here, set $m=m_{4k}$. Let $\mathbf R$ be the physical position in cubic Cartesian axes and $\mathbf X$ its fractional crystallographic coordinates. Following the MnGe notation, define the radian coordinates and the physical section by
\[
\begin{aligned}
\mathbf X&=(X,Y,Z)=\mathbf R/a,\\
\tilde{\mathbf r}&=(\tilde x,\tilde y,\tilde z)=2\pi k\,\mathbf X,\\
2\pi x_{\nu+3}&=2\pi\mathbf k_\nu\cdot\mathbf X
=(\mathbf k_\nu/k)\cdot\tilde{\mathbf r}.
\end{aligned}
\]
Here $\nu=1,\ldots,4$ and $k$ is the common positive cubic component of the propagation vectors in reciprocal-lattice units. At a Mn site, $\mathbf X=\mathbf n+\mathbf r_j$, with integer cell translation $\mathbf n$ and fractional basis position $\mathbf r_j$. The internal coordinates $x_{\nu+3}$ and the common phase $\alpha_0$ are in cycles; the tilded coordinates include the factor $2\pi$. Let $c_u=\cos u$, $s_u=\sin u$ for $u=\tilde x,\tilde y,\tilde z$.
Introduce $C_0=\cos(2\pi\alpha_0)$, $S_0=\sin(2\pi\alpha_0)$,
$C_\pm=\cos[2\pi(\alpha_0\pm1/3)]$ and
$S_\pm=\sin[2\pi(\alpha_0\pm1/3)]$.
Trigonometric reduction then gives the dimensionless four-arm field $\mathbf m=\mathbf M/m_{4k}$:
\begin{equation}
\mathbf m=-4
\begin{pmatrix}
C_0c_{\tilde x}s_{\tilde y}s_{\tilde z}+S_0s_{\tilde x}c_{\tilde y}c_{\tilde z}\\
C_+s_{\tilde x}c_{\tilde y}s_{\tilde z}+S_+c_{\tilde x}s_{\tilde y}c_{\tilde z}\\
C_-s_{\tilde x}s_{\tilde y}c_{\tilde z}+S_-c_{\tilde x}c_{\tilde y}s_{\tilde z}
\end{pmatrix}.
\label{eq:continuum_field}
\end{equation}
The field $\mathbf M=m_{4k}\mathbf m$ reproduces all four Mn sublattices for arbitrary $\alpha_0$, retaining the spatial variation of the moment magnitude. Analogous continuous fields for the two-arm ($3+2$D) and three-arm ($3+3$D) configurations follow by retaining, respectively, two or three arm contributions in Eq.~\ref{eq:common_modulation}, making the same phase-coordinate substitution, and simplifying trigonometrically. The common-field condition and explicit one-, two- and three-arm expressions are detailed in Sec.~S9.

For the four-arm 3+4D structure, only one free phase remains after Eq.~\ref{eq:ideal_helix_polar}; it is added equally to all four arms. In the continuous $3+3$D MnGe model~\cite{PomjakushinMnGe2023}, the three propagation vectors along the Cartesian axes are linearly independent. The overall phases of the three helices can therefore be absorbed into a spatial translation, moving the zeros without changing their charges. For four-arm MnSi, however, $\sum_\nu\mathbf k_\nu=0$, so a spatial translation leaves the sum of the arm phases unchanged, whereas the common phase contributes $4\alpha_0$ to this sum modulo one. An arbitrary common phase thus cannot be removed by translation and can change the cancellation pattern and its zeros, although all first-order magnetic satellite intensities remain unchanged. Phase-dependent topology has also been analysed theoretically for three- and four-arm spin textures~\cite{Shimizu2022}.

For the following calculation of zeros and topological charges we work with the same radian coordinates $\boldsymbol\xi=\tilde{\mathbf r}=(\tilde x,\tilde y,\tilde z)$ in the conventional repeat cube $0\leq\tilde x,\tilde y,\tilde z<2\pi$. Physical coordinates, the lattice constant and the modulation length are not required for these calculations. The unnormalized field is smooth and satisfies $\nabla_{\boldsymbol\xi}\cdot\mathbf m=0$. Its zeros are singularities of the direction $\hat{\mathbf m}=\mathbf m/|\mathbf m|$, whose charge on an outward-oriented sphere enclosing one isolated zero is
\begin{equation}
Q=\frac{1}{4\pi}\int
\hat{\mathbf m}\cdot
(\partial_\theta\hat{\mathbf m}\times\partial_\psi\hat{\mathbf m})
\,d\theta\,d\psi .
\label{eq:continuum_charge}
\end{equation}
Here $\theta,\psi$ parametrize the sphere in $\boldsymbol\xi$ space; a radial outward texture has $Q=+1$.

For $\alpha_0=0$, the isolated zeros occur at
\begin{equation}
\boldsymbol\xi_0=\frac{\pi}{2}(1+2n_x,1+2n_y,1+2n_z),
\label{eq:continuum_nodes}
\end{equation}
with $n_x,n_y,n_z=0,1$ in this cube. Their charges are $Q=(-1)^{n_x+n_y+n_z}$: four positive and four negative. There are also four zero-moment lines parallel to $\tilde x$, at $\tilde y=n_y\pi$, $\tilde z=n_z\pi$, which do not carry isolated-point charges.

For example, changing the common phase to $\alpha_0=0.02/(2\pi)\simeq0.00318$ lifts the zero-moment lines except at eight points, $\boldsymbol\xi=\pi(n_x,n_y,n_z)$, which become isolated zeros, while retaining the original eight isolated zeros and their charges. The cube then contains eight charges of each sign. A single ideal helix has constant nonzero magnitude and no such zeros. These results characterize particular continuous four-arm configurations. Within the present model, the magnetic Bragg intensities determine neither the common phase nor whether the four arms coexist coherently or occur in separate single-arm domains. The calculations therefore illustrate possible topologies of structures compatible with the diffraction data.

\begin{figure*}[!t]
\centering
\includegraphics[width=\textwidth]{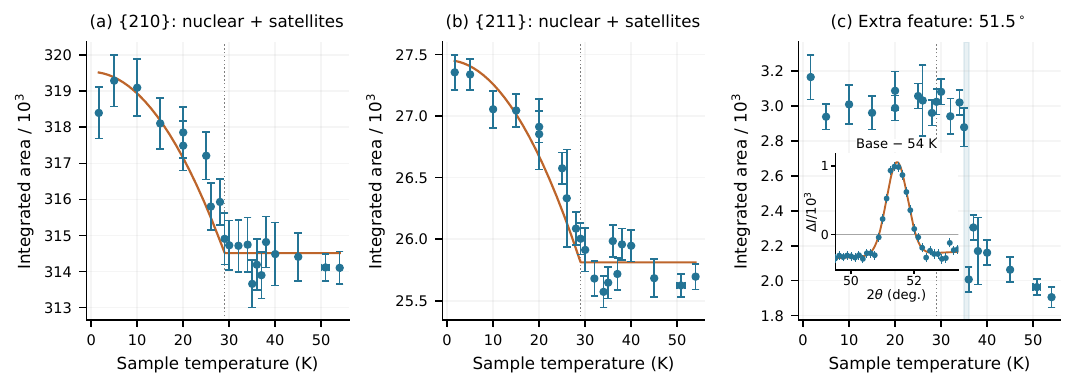}
\caption{DMC temperature scans over 1.65--54~K. The main panels use unsubtracted patterns. (a,b) Fitted total areas of nuclear peaks and unresolved magnetic satellites. Orange curves are the empirical fits described in Sec.~S8, with $T_0=29$~K fixed. (c) Direct integral of the $51.5^\circ$ feature, with a local linear background. Areas retain the common monitor normalization and angular integration step.  Vertical bars include counting and empirical uncertainty contributions; horizontal bars show sampled-temperature RMS spread. The dotted line marks 29~K; the shaded interval in (c) brackets the observed extra-band decrease. Inset in (c): base-minus-54~K difference from the final DMC campaign; the solid curve is a Gaussian peak fit with a linear background.}
\label{fig:DMC_T_final}
\end{figure*}

\section{Temperature-dependent intensities and additional scattering}
\label{sec:T_dependence}

An additional positive contribution is observed in some base-minus-warm patterns from both HRPT and DMC. It lies near $Q\simeq2.22$~\AA$^{-1}$, corresponding to $2\theta\simeq51.4^\circ$--$51.5^\circ$ at $\lambda\simeq2.45$~\AA. The inset in Fig.~\ref{fig:DMC_T_final}(c) illustrates its resolved peak profile. In the examined base-minus-warm comparisons, resolved negative contributions near $43.2^\circ$ and $46.2^\circ$ on the flanks of the strong $\{110\}$ group occur together with this positive feature. These signed differences are not directly identified with separate magnetic reflections of MnSi.

To reduce dependence on a single subtracted reference, we also examine integrated areas in the unsubtracted DMC patterns. Nuclear peaks and unresolved magnetic satellites are integrated together, with a local background. Integrating each complete group reduces sensitivity to small peak shifts and redistribution of intensity within the group. The feature near $51.5^\circ$ is evaluated over a separate band, which also contains overlapping nuclear scattering.  Figure~\ref{fig:DMC_T_final} shows the integrated intensities as a function of temperature, demonstrating the decrease in the MnSi magnetic satellites above $T_N$.

Between 25 and 30~K, the total $\{210\}$ and $\{211\}$ areas decrease by $0.78(30)\%$ and $2.49(81)\%$, respectively, while the additional $51.5^\circ$ band changes by only approximately $(0.8\pm3.3)\%$; no significant change of this band is resolved between 25 and 30~K. The main-peak changes are consistent with loss of the MnSi magnetic contribution near 29~K. 
The extra band instead decreases between 35 and 36~K by about 30\% of its 35~K area for the stated linear background. The contrasting trends suggest that the additional scattering does not simply follow the principal MnSi magnetic intensity.

An unusual aspect is the dependence on measurement history. The positive $51.4^\circ$--$51.5^\circ$ temperature-difference feature is prominent in some measurement campaigns and strongly suppressed in others, while impurity-related nuclear scattering remains visible in the unsubtracted patterns. Its suppression therefore does not imply the disappearance of a crystalline minority phase. 
The origin of this measurement-history dependence remains unconfirmed.

The two-dimensional DMC data distinguish the positive feature from the negative flanks. The latter can be concentrated within a limited range of detector heights, whereas the strong positive difference extends over most of the recorded height. These distinct spatial signatures do not establish a common origin for all three features. The spatial signatures alone, however, cannot distinguish a minority-phase contribution from changes in the host or reference profiles. 

An interesting observation made late in this study provides an indirect test of the impurity interpretation. A subsequent HRPT dataset (H4), recorded a few weeks after the DMC measurements (Fig.~\ref{fig:DMC_T_final}) and after the powder container had been shaken, shows a strongly reduced extra temperature-difference feature, with local profile fits favouring a weak residual, while its ideal-helix amplitude agrees with H1--H3 (Sec.~S4\,C and Fig.~S6). This contrast favours a magnetic minority-phase contribution from a few coarse grains or a mosaic crystallite whose diffraction contribution depends on orientation. An oxide phase distinct from MnSi is a plausible candidate, with surface oxides on the Mn starting material before synthesis providing a possible source.

Although this additional scattering is peripheral to the MnSi magnetic-structure determination, it was present during the extended measurements underlying the temperature-dependent analysis. We therefore report its behaviour for completeness. Its origin remains unresolved, and we exclude these features from the MnSi magnetic-structure interpretation. The general profile and subtraction checks are described in Sec.~S2.

\section{Discussion}

MnSi illustrates a general difficulty in long-period magnets: establishing the propagation vector and helicity does not fully specify how the modulation is distributed over the crystallographic basis; a single small-angle Bragg intensity is insufficient for this purpose. In the present case, the observation of small-angle magnetic scattering in numerous experiments nevertheless excludes the pure mLD1 model, whose Fourier component at the propagation vector is longitudinal, leaving mLD2 and its complex-conjugate partner mLD3. Wide-angle satellite intensities provide the additional information needed to constrain the modulation over the crystallographic basis. As discussed in the Introduction, early neutron studies established key features of the helical order~\cite{Ishikawa1976,Ishida1985}, but did not refine the full site-resolved modulation against zero-field satellite intensities. The most revealing structural clues, however, need not lie in the most prominent peaks. In retrospect, a sensitive test of the simultaneous extinction of the two $\{100\}$ satellite groups could already have placed stringent constraints on the modulation and provided a quantitative test of the ideal-helix interpretation.

The zero-field SANS map of Ishikawa et al.~\cite{Ishikawa1976} shows four fundamental spots in the measured plane, identifying propagation along $\langle111\rangle$ with a period of approximately 180~\AA. This period characterizes the single helix or each constituent helix of the four-arm superposition, so both descriptions are compatible with these SANS observations. 

Experimental difficulty is one reason why a systematic wide-angle intensity refinement of the zero-field magnetic structure has remained elusive despite the extensive literature on MnSi. As quantified in Sec.~\ref{mag_str}, long counting times alone are insufficient: extracting the weak satellites requires reproducible nuclear peak positions, widths and intensities, while the nonmonotonic lattice response further complicates subtraction.

Interpreting these intensities also requires a sufficiently general, but manageable, description of the allowed magnetic structures. A methodological contribution of this work is the explicit non-standard cubic $3+4$D construction, which provides a common description of four-arm configurations and their three- and two-arm restrictions. Retaining a separate phase coordinate for each arm makes the cubic component and intersite phase relations explicit across the family of phase sections. These relations constrain the first-harmonic Mn modulation to eight real coefficients and remain imposed when only two or three arms are retained. By comparison, the conventional four-arm $3+3$D description retains all four helices at a fixed phase sum but, for the order parameter considered in this case, has an orthorhombic MSSG permitting 24 real coefficients. The lower-symmetry two- and three-arm MSSGs likewise permit 24. Our construction therefore selects a constrained family within these broader models; its smaller parameter count follows from the retained cubic relations. This common description provides a systematic basis for comparing configurations and testing departures from the ideal-helix limit supported by the diffraction data.

Such tests are relevant to microscopic models of the magnetic interactions. Chizhikov and Dmitrienko showed that a nearest-neighbour exchange and Dzyaloshinskii--Moriya model of MnSi-type crystals allows twisting and canting within the crystallographic unit cell in addition to the long-period rotation~\cite{ChizhikovDmitrienko2012}. In their bond convention and to leading order, the helix wave number depends on $D_x-2D_y-D_z$, whereas a distinct combination, $D_x+D_z$, controls a residual canting that persists even after the helix is unwound by a magnetic field. Different microscopic interaction parameters can therefore reproduce the same helix period while predicting different relative orientations of the Mn moments within the crystallographic unit cell. Related calculations incorporating several neighbour shells and more general symmetry-allowed interactions likewise distinguish the long-period rotation from relative phase shifts and canting between the magnetic sublattices~\cite{ChizhikovDmitrienko2013,HallCurnoe2021}. The wide-angle satellite intensities provide an additional experimental test of these predicted arrangements.

The same distinction matters for the connection between microscopic and continuum descriptions. A long modulation period permits a separation of length scales, but does not by itself require all crystallographic sublattices to sample the same smooth magnetization field. In our construction, that common-field description follows only for specific relations between the site-dependent Fourier coefficients. The ideal-helix constraints satisfy these relations and account for the observed intensity distribution with a single magnetic amplitude. Their agreement with independent datasets therefore supports this description within the experimental sensitivity.

The crystal metrics provide a complementary test of the magnetic configuration through magnetoelastic coupling~\cite{PlumerWalker1982}. A single-domain ideal helix and an equal-amplitude superposition of four tetrahedral arms can have the same mean-square moment but different spatially averaged tensors $\langle M_iM_j\rangle$: the former is anisotropic, whereas the latter is isotropic at this quadratic level. These configurations can consequently favour different symmetry-breaking strains, even when their powder magnetic profiles coincide. However, an equal population of single-arm domains also gives an isotropic domain-averaged tensor; distinguishing these possibilities requires sensitivity to the distortions within individual domains, for example through peak splitting or broadening. The nonmonotonic cubic lattice parameter characterizes the volume response but does not, by itself, select the number of coherently coexisting arms. Likewise, the approximately $10^{-5}$ precision of the mean lattice parameter is not a direct detection limit for rhombohedral strain. A quantitative comparison requires the relevant magnetoelastic couplings, elastic stiffnesses and experimental line-shape sensitivity.

Magnetic symmetry also matters for electronic topology: Wilde \textit{et al.} demonstrated symmetry-enforced topological nodal planes in the electronic band structure of field-polarized MnSi~\cite{Wilde2021}. Their result concerns momentum-space topology in a field-polarized state, whereas our continuum calculations address real-space magnetic textures in zero field. 

Finally, determining the modulation on the crystallographic basis and determining the coherent real-space texture are related but distinct tasks. In the finite-field skyrmion phase of MnSi, mixed higher-order reflections have been used to constrain phase relationships between helices~\cite{Adams2011}, while measurements combining neutron scattering and microwave spectroscopy have supplied complementary information through the collective excitations~\cite{Kindervater2019}. These studies illustrate how higher-order scattering and collective excitations can complement the fundamental satellite intensities. Our zero-field powder measurements constrain the modulation associated with each active arm, but do not uniquely distinguish single-arm domains from a coherent four-arm superposition or determine its common phase. The calculated zeros and topological charges therefore characterize particular configurations compatible with the diffraction results. The symmetry-constrained structures established here provide concrete candidates for further tests through their magnetic excitations, magnetoelastic response and phase-sensitive higher-order scattering.

\section{Conclusions}
\label{sec:conclusions}

Wide-angle neutron powder diffraction and magnetic-superspace symmetry analysis provide a quantitative description of the zero-field magnetic structure in MnSi associated with the four-arm $\{ k k k\}$ propagation-vector star. The analysis covers 20 independent satellite groups over the range $Q\simeq0.5$--$3.8$~$\text{\AA}^{-1}$. The independent diffraction datasets yield consistent relative intensities and modulation amplitudes.

The mLD2 representation results in a single-arm $3+1$D description with MSSG $R3.1'(0,0,g)ts$ and the explicit non-standard cubic four-arm $3+4$D construction $P2_1 3\,1'(a,a,a)\allowbreak\,0{-t}s\allowbreak (a,-a,-a)\,\allowbreak 0{-t}s\allowbreak (-a,a,-a)\,0{-t}s\allowbreak (-a,-a,a)\,0{-t}s$. Both descriptions have eight symmetry-allowed real magnetic parameters. The lower-symmetry $3+2$D and $3+3$D MSSGs considered here permit 24 coefficients. Nevertheless, the experimental datasets are very well described by the ideal transverse circular-helix limit, with a single adjustable magnetic amplitude. In this limit, the single-arm and four-arm solutions give identical powder profiles despite their different real-space moment distributions. The single-arm moment has a constant magnitude of $0.535(3)\,\mu_{\rm B}$ per Mn. The illustrated four-arm configuration has the same root-mean-square moment, with local magnitudes ranging from zero to $0.874(5)\,\mu_{\rm B}$.

The crystal structure refinements are consistent with near-stoichiometric cubic MnSi, with no resolved symmetry-lowering distortion. The mean lattice parameter exhibits a nonmonotonic temperature dependence and is determined with a formal relative precision better than $10^{-5}$. The retained cubic metric provides a structural argument in favour of the cubic $3+4$D description among models giving comparable diffraction fits.

\section*{A{\lowercase{cknowledgements}}}

This work was performed at the Swiss Spallation Neutron Source SINQ, Paul Scherrer Institute. We acknowledge financial support from the Swiss National Science Foundation through Grants No.~200020\_182536/1, No.~200021\_188706, and from the Swiss Nanoscience Institute.

V.P. is grateful to B.~J.~Campbell for valuable discussions and suggestions concerning the construction of the cubic $3+4$D magnetic superspace group. We also thank J.~M.~Perez-Mato for stimulating discussions on superspace symmetry and magnetic irreducible representations. We thank D. Sheptyakov for the help with DMC experiments.

 \bibliography{}

\end{document}


\title{Supplementary Material for ``Crystal and magnetic structures of MnSi
in zero magnetic field: Neutron diffraction and symmetry analysis''}
\author{V.~Pomjakushin}
\affiliation{Laboratory for Neutron Scattering and Imaging, Paul Scherrer
Institut, CH-5232 Villigen PSI, Switzerland
}
\author{I.~Plokhikh}
\affiliation{Laboratory for Multiscale Materials Experiments, Paul Scherrer Institut, CH$-$5232 Villigen PSI, Switzerland
}
\affiliation{Physics Department, Technical University Dortmund, Otto-Hahn-Stra{\ss}e 4, 44227 Dortmund, Germany}
\affiliation{Research Center Future Energy Materials and Systems, 44227 Dortmund, Germany}
\author{ D.~H.~Fabini}
\affiliation{Laboratory for Neutron Scattering and Imaging, Paul Scherrer
Institut, CH$-$5232 Villigen PSI, Switzerland
}

\author{E.~Pomjakushina}
\affiliation{Laboratory for Multiscale Materials Experiments, Paul Scherrer Institut, CH-5232 Villigen PSI, Switzerland}
\maketitle

In this Supplementary Material (SM), section, equation, figure and table
numbers carry the prefix S. Unprefixed numbers refer to the main text.

\section{Experimental datasets and preparation}
\label{sm:experimental}
The datasets below support the experimental description in Sec.~II
of the main text. The H0--H4 labels identify independent HRPT datasets
and are used consistently throughout the analysis.

\subsection{HRPT magnetic-diffraction datasets}
\label{sm:HRPT_datasets}
H1, H2 and H3 were measured in HRPT high-intensity mode at
$\lambda\simeq2.45$~\AA. Each consists of a base-temperature pattern
and a 54~K reference pattern. Table~\ref{tab:SM_exposures} lists the
counting exposures, together with the earlier H0 and later H4 datasets
described below.
The total for H1--H3 is approximately 360~h, including both the
base-temperature and reference measurements.

\begin{table}[htbp]
\caption{Approximate counting exposures of the HRPT magnetic datasets.
The reference temperature
is 40~K for H0 and 54~K for H1--H4. H0 and H4 are additional to the
approximately 360~h used for the H1--H3 comparison.}
\label{tab:SM_exposures}
\centering
\begin{ruledtabular}
\begin{tabular}{llrr}
Dataset & Measurement date & Base temperature (h) & Warm reference (h)\\
H0 & November 2022 & 36.0 & 37.0\\
H1 & May 2023 & 41.8 & 26.7\\
H2 & July 2025 & 41.0 & 40.6\\
H3 & October 2025 & 118.7 & 89.9\\
H4 & October 2026 & 35.8 & 28.2\\
\end{tabular}
\end{ruledtabular}
\end{table}

The earlier HRPT dataset H0 was measured at $\lambda=1.886$~\AA\
with mean sample temperatures of 1.5 and 40~K. Its 
base-minus-40~K difference provides an additional observation of the
extra feature, discussed in Sec.~\ref{sm:H0}. It is analysed using
parameters from its own 40~K nuclear refinement and is not included
in the H1--H3 comparisons or averages.

The later HRPT dataset H4 was completed in October 2026 in high-intensity
mode at $\lambda=2.450$~\AA, with mean sample temperatures of 1.5 and
54~K. H4 is analysed separately in Sec.~\ref{sm:H4}; it is not
included in the H1--H3 averages used in the main text.

\subsection{HRPT structural datasets}
\label{sm:structural_datasets}
The $\lambda=1.155$~\AA\ structural temperature scan comprises
110 independently refined patterns. Individual patterns are retained
for the lattice-parameter analysis, with 1~K temperature-bin averages
used for presentation. In contrast, the patterns measured over 2--6~K
were merged for the low-temperature crystal-structure refinement.
Merging these diffraction patterns is distinct from averaging the
independently refined lattice parameters.

The complementary HR-mode patterns were measured at
$\lambda=1.494$~\AA\ and recorded mean temperatures of 1.8 and
55~K. The 55~K HR pattern used in the structural comparison is distinct
from the 54~K HI reference patterns used for magnetic subtraction.
The HR instrumental line shape was calibrated using
Na$_2$Al$_2$Ca$_3$F$_{14}$ (NAC).

\subsection{DMC datasets}
\label{sm:DMC_datasets}
The DMC base-minus-54~K dataset was measured at
$\lambda\simeq2.45$~\AA\ and retains its native $0.125^\circ$
spacing in $2\theta$. It supplies the sensitive check of the first
$\{100\}$ satellite groups. The separate temperature-scan dataset
is used for integrated intensities in the unsubtracted patterns;
it is not pooled with the extinction-check dataset.
 Integration, averaging and scan-history comparisons
are described in Sec.~\ref{sm:temperature}.

The DMC/HRPT count-rate comparison uses background-subtracted areas
of strong MnSi nuclear peaks from the October 2026 raw detector data,
normalized by counting time and proton-monitor current. The integration
accounts for the different angular bin widths. The full-detector ratio is approximately 50--60 for the strong peaks used in this comparison. It includes detector acceptance and efficiency. Correcting for the approximately 2.5-fold larger vertical angular acceptance of DMC gives an estimated incident neutron-flux ratio at the sample of about 20--25, assuming comparable detector efficiencies, illuminated sample volumes and transmission.

\begin{table}[htbp]
\caption{DMC datasets compared in Fig.~\ref{fig:SM_DMC_three_differences}.
``Warm'' denotes a 54~K reference.}
\label{tab:SM_DMC_campaigns}
\centering
\begin{ruledtabular}
\begin{tabular}{lll}
Curve & Measurement date & Difference definition\\
July original selection & July 2026 & Pooled base $-$ pooled warm\\
October base 1 & October 2026 & First base $-$ first warm\\
October base 2, warm 1 & October 2026 & Second base $-$ first warm\\
October base 2, warm 2 & October 2026 & Second base $-$ second warm\\
\end{tabular}
\end{ruledtabular}
\end{table}

\section{DMC profile diagnostics and subtraction limitations}
\label{sm:DMC_diagnostics}
The base-minus-54~K DMC pattern was checked independently of the
ideal-helix intensity constraints. FullProf supplied the indexed
satellite profiles with fixed angular calibration, lattice parameter,
propagation vector and line-shape parameters. The amplitudes of the
18 satellite groups around $\{110\}$, $\{111\}$, $\{200\}$,
$\{210\}$ and $\{211\}$ were then varied independently by weighted
least squares. The two $\{100\}$ groups were set to zero, but their
measured bins were retained. The original $0.125^\circ$ grid and
counting uncertainties were used over $20^\circ$--$88^\circ$,
excluding $50^\circ$--$53^\circ$ around the extra feature; the
FullProf grid mask retains $N=518$ data points.

For the weighted fits in this section, we report
\begin{equation}
\chi^2=\frac{S}{N-n_{\rm par}},\qquad
S=\sum_{i=1}^{N}\left[\frac{\Delta I_i-I_{{\rm calc},i}}
{\sigma_{\Delta I,i}}\right]^2,
\label{eq:SM_DMC_chi2}
\end{equation}
where $n_{\rm par}$ counts the varied intensity and background
coefficients. With nonnegative intensities and the saved background
fixed, $S=3473.04$ and $\chi^2=6.9461$ ($n_{\rm par}=18$).
Adding a constant, linear or quadratic background correction gives
$\chi^2=6.5530$, 6.5650 and 6.5754, respectively
($n_{\rm par}=19$, 20 and 21). These are nominal parameter-count
statistics for the constrained fits. Allowing signed intensities
reduces the fixed-background value only to $\chi^2=6.6577$ and
produces three negative coefficients.

\begin{figure}[htbp]
\centering
\includegraphics[width=\textwidth]{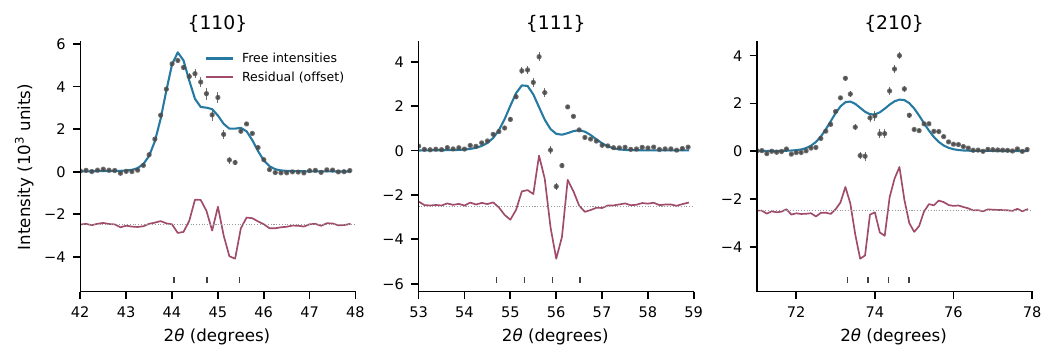}
\caption{DMC base-minus-54~K profile diagnostic. Points retain the
propagated counting-error bars; blue curves use 18 independently
varied nonnegative satellite-group intensities. The 
background is removed from both the data points and calculation.
Purple curves are their differences, offset downwards; short ticks
mark the indexed satellite positions. Shape residuals persist without
ideal-helix intensity constraints.}
\label{fig:SM_DMC_diagnostic}
\end{figure}

Figure~\ref{fig:SM_DMC_diagnostic} shows that structured residuals remain at the strong satellite groups even when their relative intensities are free. Smooth background corrections do not resolve this mismatch. Thus the residuals cannot by themselves establish a departure from the ideal-helix magnetic structure. The diagnostic highlights the limitations of nuclear subtraction here: small differences in the data point intensities of the much stronger nuclear peaks leave residuals comparable to the weak magnetic signal.

\begin{figure}[htbp]
\centering
\includegraphics[width=\textwidth]{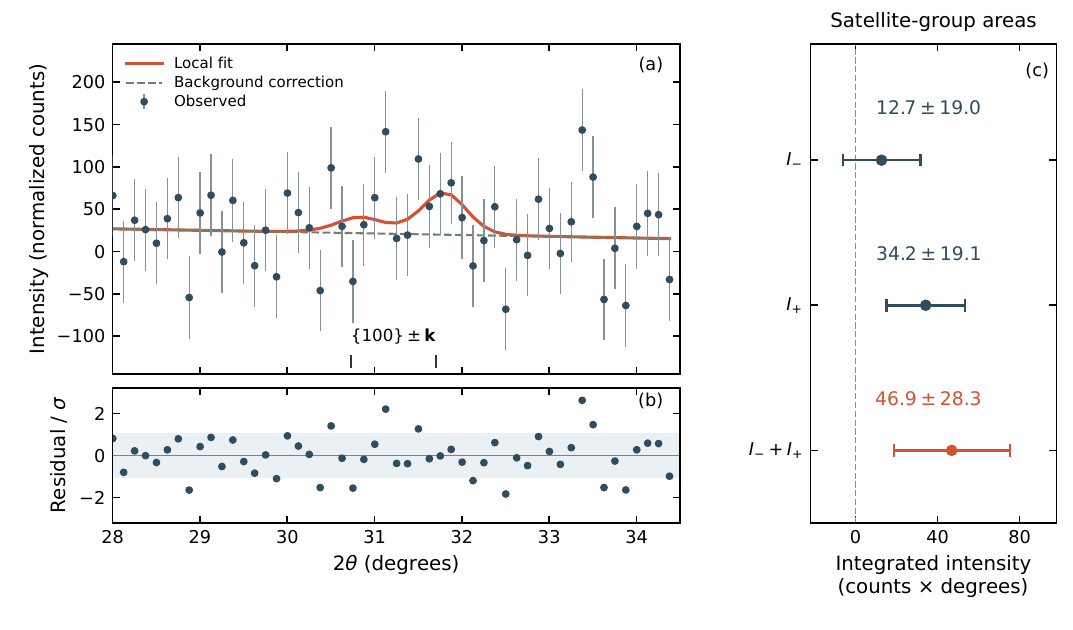}
\caption{Local DMC test for the two $\{100\}$ satellite groups.
(a) Base-minus-54~K data points with error bars and the fitted profile (red). The fixed
background is removed from both curves; its fitted linear correction
is shown separately (dashed). Short ticks mark the two indexed group
positions. Their two areas are fitted without a positivity constraint,
with positions and widths fixed. (b) Residuals divided by the original counting
uncertainties; shading marks $\pm1$ as a visual guide. The fit uses
52 native-grid bins and four parameters, giving $\chi^2=0.960$.
(c) Integrated areas of the lower-angle ($I_-$) and higher-angle ($I_+$)
groups and their sum, with one-standard-deviation error bars.
The values are $12.7\pm19.0$, $34.2\pm19.1$ and $46.9\pm28.3$,
respectively, in the same normalized intensity-times-degree units.
The sum error includes the inter-group covariance; the total lies only
$1.66\sigma$ above zero.}
\label{fig:SM_DMC_100_local}
\end{figure}

The DMC observation of absent resolved $\{100\}$ satellites is a
separate local test: fitting their two indexed profiles and a linear
background correction over $28^\circ$--$34.5^\circ$ gives
$\chi^2=0.960$ for 52 bins and four parameters
(Fig.~\ref{fig:SM_DMC_100_local}). 

The separate native DMC Le Bail refinement gives the cubic propagation
component $k=0.01510(10)$. The native Le Bail statistic listed
there is $S/N=8.2918$, distinct from the direct weighted diagnostic above.
The value of $k$ is held fixed in the subsequent magnetic calculations.

\begin{figure}[!htbp]
\centering
\includegraphics[width=\textwidth]{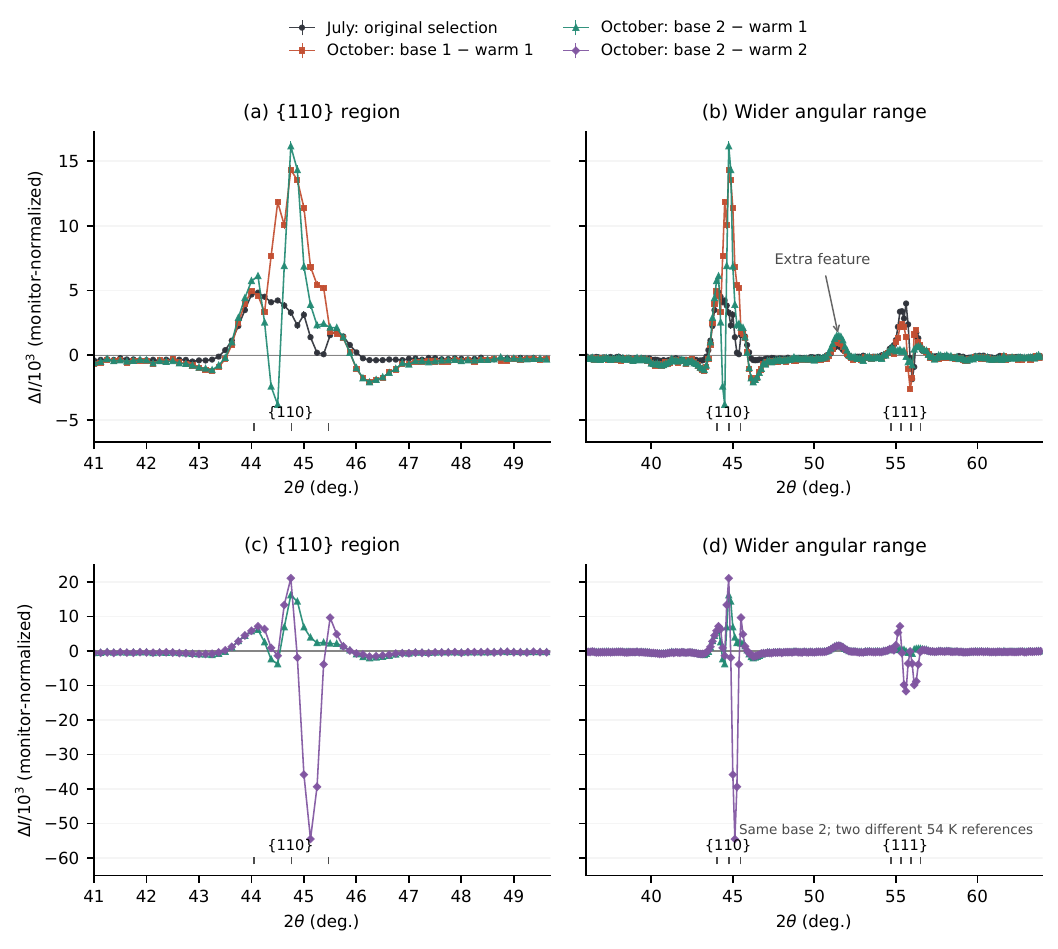}
\caption{DMC base-minus-54~K profiles from July and October 2026.
(a,b) The original pooled July selection and the first and second
October base measurements minus the first 54~K reference (``warm 1'').
(c,d) The same second base measurement minus warm 1 or the completed
second reference (``warm 2'', runs 10536--10611); the vertical range
is enlarged to show the latter subtraction. Warm 2 pools the complete
measurement, including its observed profile change. The left panels
show $\{110\}$; the right panels include $\{111\}$ and the additional
feature near $51.5^\circ$. All curves retain the common $10^8$ monitor
normalization, $0.125^\circ$ bins and original angular positions and
baselines, without fitted relative scales or vertical offsets.
Bars show propagated one-standard-deviation counting errors; curves
sharing a base or reference have correlated errors. Connecting lines
guide the eye; short ticks mark satellite positions from the July
DMC model.}
\label{fig:SM_DMC_three_differences}
\end{figure}

The reproducibility of the temperature differences is illustrated in
Fig.~\ref{fig:SM_DMC_three_differences}. The original July profile
and two October profiles using the first 54~K reference are retained
in panels (a,b). For the latter pair, subtracting the two curves
cancels the shared reference and isolates the difference between the
base-temperature measurements. Panels (c,d) use the same second
base-temperature profile with either of the two 54~K references,
thereby isolating the influence of the reference choice. The completed
second 54~K reference includes both sides of a profile change within
that measurement, already apparent in comparisons of its individual
counting blocks. The larger residuals obtained with this
pooled reference illustrate why these differences cannot be interpreted
as magnetic scattering alone. These comparisons do not identify the
cause of the profile changes.

\clearpage

\section{Crystal-structure refinement and precision checks}
\label{sm:crystal_checks}
\subsection{High-resolution profiles and mean-cell precision}
The HR refinements at $\lambda=1.494$~\AA\ cover
$11^\circ<2\theta<164^\circ$ with 32 varied parameters, including
18 background coefficients (Table~\ref{tab:SM_HR_checks}).
Parameter errors are formal FullProf uncertainties and exclude
uncertainties in the fixed wavelength and instrumental parameters.

\begin{table}[htbp]
\caption{HR crystal-profile checks. $U$ is a correction to the
instrumental Gaussian coefficient, not the total width coefficient.
$\chi^2=S/(N-n_{\rm par})$ follows Eq.~\eqref{eq:SM_DMC_chi2}.}
\label{tab:SM_HR_checks}
\centering
\begin{ruledtabular}
\begin{tabular}{rrrrr}
$T$ (K) & $a$ (\AA) & $U$ (deg$^2$) & $R_{\rm wp}$ (\%) & $\chi^2$\\
1.8 & 4.552745(25) & $-0.00382(35)$ & 5.66 & 3.3254\\
55 & 4.552748(25) & $-0.00401(34)$ & 5.53 & 3.1950\\
\end{tabular}
\end{ruledtabular}
\end{table}

The Gaussian squared FWHM is
\begin{equation}
H_G^2=(U_{\rm inst}+U)\tan^2\theta
 +(V_{\rm inst}+V)\tan\theta+W_{\rm inst}+W,
\label{eq:SM_width}
\end{equation}
where the NAC coefficients are
$(U_{\rm inst},V_{\rm inst},W_{\rm inst})=
(0.027891,-0.083288,0.101028)$~deg$^2$.
The instrumental Lorentzian component is retained; $U,V,W$ are
refined corrections, so negative $U$ does not represent negative
strain. The minimum instrumental FWHM corresponds to
$\Delta d/d\simeq1.05\times10^{-3}$, whereas
$\sigma_a/a\simeq5.5\times10^{-6}$ describes mean-cell precision,
not a peak-splitting limit.

For an illustrative Gaussian rms strain $\epsilon$,
$\Delta U=32\ln2\,(180/\pi)^2\epsilon^2$. Taking
$\sigma_U\simeq3.4\times10^{-4}$~deg$^2$ gives
\begin{equation}
\epsilon_{\rm sens}=\frac{\pi}{180}
\sqrt{\frac{\sigma_U}{32\ln2}}\simeq6.9\times10^{-5}.
\label{eq:SM_strain_sensitivity}
\end{equation}
This is not a confidence bound on rhombohedral strain: the fitted
correction is negative, width parameters are correlated, and no
symmetry-lowered profile was fitted. The temperature change
$U(1.8\,{\rm K})-U(55\,{\rm K})=0.00019(49)$~deg$^2$
is consistent with zero using independent counting errors.

\subsection{Full-range occupancy comparison}
The matched $\lambda=1.155$~\AA\ fits use 3019 data points over
$11^\circ<2\theta<162^\circ$ ($Q_{\max}\simeq10.75$~\AA$^{-1}$)
and the same background and profile model. The Si site
remains fully occupied when the Mn population is varied, removing the
common scale--occupancy ambiguity.

\begin{table}[htbp]
\caption{Matched full-range occupancy refinements of the merged 2--6~K
HI pattern. The isotropic atomic displacement parameters are given as
$B=8\pi^2U_{\rm iso}$. Errors are formal FullProf errors.}
\label{tab:SM_occupancy}
\centering
\begin{ruledtabular}
\begin{tabular}{lrrrrrr}
Model & Varied relative population & $B_{\rm Mn}$ (\AA$^2$) & $B_{\rm Si}$ (\AA$^2$) & $n_{\rm par}$ & $R_{\rm wp}$ (\%) & $\chi^2$\\
Full sites & --- & 0.103(9) & 0.094(8) & 32 & 4.32 & 3.9281\\
Mn varied, Si fixed & $g_{\rm Mn}=0.965(14)$ & 0.065(17) & 0.127(15) & 33 & 4.31 & 3.9247\\
\end{tabular}
\end{ruledtabular}
\end{table}

Varying the Mn population gives only
a small improvement in $\chi^2$. Correlated profile residuals persist;
the small improvement and formal occupancy error do not establish
a chemical deviation from stoichiometry.

\section{HRPT profile decomposition and dataset comparison}
\label{sm:weighted}
\subsection{Calculated profiles, fitted intensities and uncertainties}
For H1--H3 and H4, FullProf supplies satellite profiles with the lattice,
wavelength, angular, width and asymmetry parameters and nuclear scale
fixed to each dataset's 54~K refinement; $k$ is fixed to its ideal-helix
estimate. The nuclear scale is the Rietveld intensity factor with common
structural normalization, not an individual peak area.
After fixing the two $\{100\}$ intensities to zero, 296 reflection entries
form 18 groups of coincident reflections, each with one freely fitted
combined intensity. The complete $\{110\}$ family comprises three such
groups. The measured $\{100\}$ bins remain included; the independent
local extinction test is in Sec.~\ref{sm:DMC_diagnostics}.

With $P_g(2\theta_i)$ the fixed profile of group $g$ at measured bin $i$,
\begin{equation}
 I_i^{\rm calc}=\sum_{g=1}^{18}c_g P_g(2\theta_i)
                 +b_0+b_1t_i+b_2t_i^2,\qquad
 t_i=\frac{2\theta_i}{80^\circ}-1.
 \label{eq:sm_linear_model}
\end{equation}
The 18 intensities and one global quadratic background are fitted
simultaneously by minimizing
$S=\sum_i[(I_i^{\rm obs}-I_i^{\rm calc})/\sigma_i]^2$.
Overlapping profiles contribute to the same bins; the ideal-helix fit
uses the same background form with independently fitted coefficients.
Native Le Bail extraction instead partitions intensity iteratively
between overlapping reflections, alternating with background and
profile refinement. Neither procedure guarantees an exactly zero
estimate for an absent peak; its uncertainty must be considered.

Displayed profiles use nonnegative coefficients; area comparisons use
unrestricted estimates to avoid truncating noise at zero. These solutions
coincide for H1--H3, whose coefficients are all positive. With $X$
containing the 18 profile and three background columns,
\begin{equation}
 \widehat{\boldsymbol\beta}=(X^TWX)^{-1}X^TW\mathbf I^{\rm obs},
 \qquad C_\beta=(X^TWX)^{-1},\qquad W_{ii}=\sigma_i^{-2}.
 \label{eq:sm_covariance}
\end{equation}
The covariance includes peak--background correlations and uses propagated
counting variances without scaling by $\chi^2$. For nuclear-scale-normalized
areas, $\mathbf A=L\widehat{\boldsymbol\beta}$ and
$C_A=LC_\beta L^T$, where $L$ includes the full profile integrals and
inverse nuclear scale.
Background uncertainty is therefore already
included; uncertainties in the fixed lattice, wavelength, angular
and profile parameters, nuclear scale and $k$ are not propagated.

We report $\chi^2=S/(N-n_{\rm par})$. The ideal fit varies amplitude,
$k$ and three background coefficients ($n_{\rm par}=5$). The free fit
has 21 linear coefficients.
Native Le Bail statistics count extracted intensities differently,
so residual sums $S$ provide the direct comparison.

\begin{table}[htbp]
\caption{Independent HRPT weighted fits. All rows use
$11^\circ<2\theta<96^\circ$ ($N=1699$), except the explicitly restricted
$20^\circ<2\theta<96^\circ$ H3 comparison ($N=1519$) used in
main-text Table~II. The displayed nonnegative free fit varies the 18
powder-position intensities and quadratic background at the ideal-fit
$k$, with nominal $n_{\rm par}=22$.}
\label{tab:sm_profile_fits}
\centering
\begin{ruledtabular}
\begin{tabular}{lrrrr}
Dataset & $S$, ideal & $\chi^2$, ideal & $S$, free & $\chi^2$, free\\
H1 & 1810.00 & 1.0685 & 1768.77 & 1.0547\\
H2 & 1921.48 & 1.1343 & 1847.52 & 1.1017\\
H3 & 2352.46 & 1.3887 & 2279.08 & 1.3590\\
H3 ($20^\circ$--$96^\circ$) & 1617.44 & 1.0683 & 1535.62 & 1.0258\\
H4 & 1935.06 & 1.1423 & 1874.72 & 1.1179\\
\end{tabular}
\end{ruledtabular}
\end{table}

On the full grid of dataset H3, native Le Bail gives $S=2470.64$,
versus 2279.08 for direct decomposition with matching fixed
nuclear-refinement parameters and $k$. Restricting H3 to
$20^\circ$--$96^\circ$ reduces the low-angle-background contribution
and gives
$k=0.01470(12)$ and $m_{1k}=0.4372(33)\,\mu_{\rm B}$.
Main-text Table~II uses this restricted fit; the profile plots and
display average retain the full interval.

\begin{figure}[htbp]
\centering
\includegraphics[width=\textwidth]{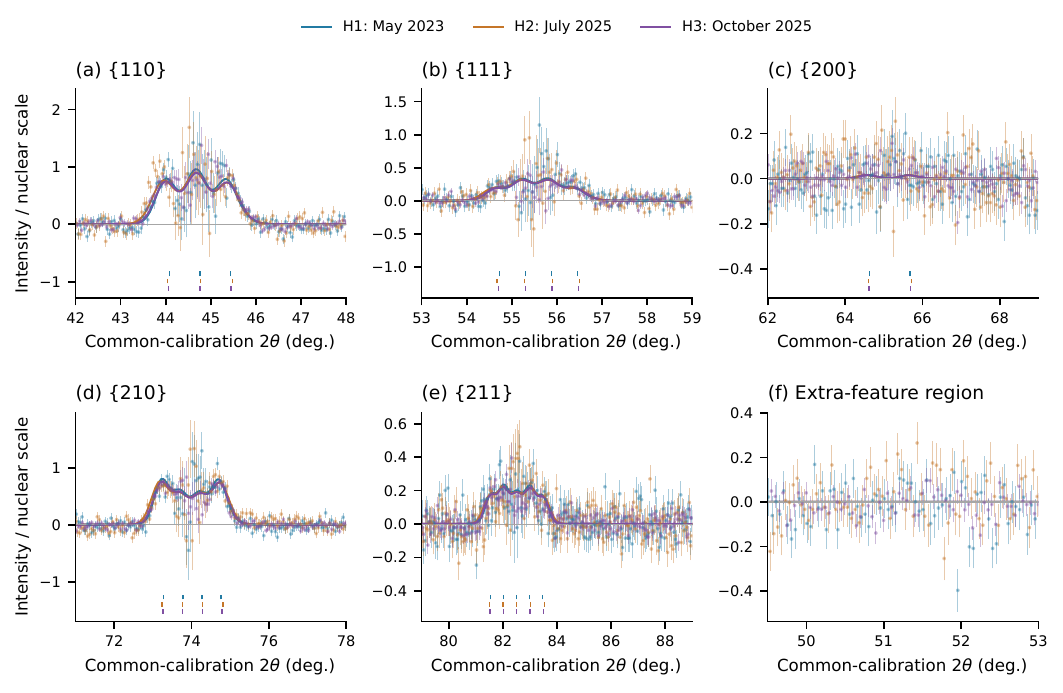}
\caption{Independent HRPT data points and ideal-helix fits for H1--H3.
Intensities and counting errors are divided by the respective 54~K
nuclear scales. Angular coordinates are transformed to a common
angular calibration for display; the original data are fitted
on their own grids. The last panel shows the extra-feature region, without
assigning a MnSi satellite to it.}
\label{fig:sm_independent_HRPT}
\end{figure}

For the main-text display average, background-subtracted intensities
are divided by their nuclear scales and mapped to a common lattice
parameter and wavelength after removing angular corrections.
Each point enters one $0.10^\circ$ bin with inverse-variance weight;
calculated curves use identical weights. The average retains the
different resolutions and $k$ values and is not refined. Its counting
errors are conditional on the background, scale and coordinate mapping.

Table~\ref{tab:sm_contrast} instead divides each complete magnetic-family
area by its corresponding calculated 54~K nuclear area in the same
units; these integrated-area ratios differ from nuclear-scale normalization.
\begin{table}[htbp]
\caption{Magnetic-to-nuclear integrated intensity ratios (\%) for the
principal allowed nuclear families. Symmetry-equivalent coincident
reflections are included in the nuclear denominator.}
\label{tab:sm_contrast}
\centering
\begin{ruledtabular}
\begin{tabular}{lrrrr}
Dataset & $\{110\}$ & $\{111\}$ & $\{210\}$ & $\{211\}$\\
H1 & 1.54 & 1.39 & 2.05 & 7.92\\
H2 & 1.60 & 1.33 & 2.04 & 8.48\\
H3 & 1.62 & 1.14 & 1.86 & 7.50\\
\end{tabular}
\end{ruledtabular}
\end{table}

\subsection{Earlier HRPT dataset H0 and the extra feature}
\label{sm:H0}
\begin{figure}[htbp]
\centering
\includegraphics[width=\textwidth]{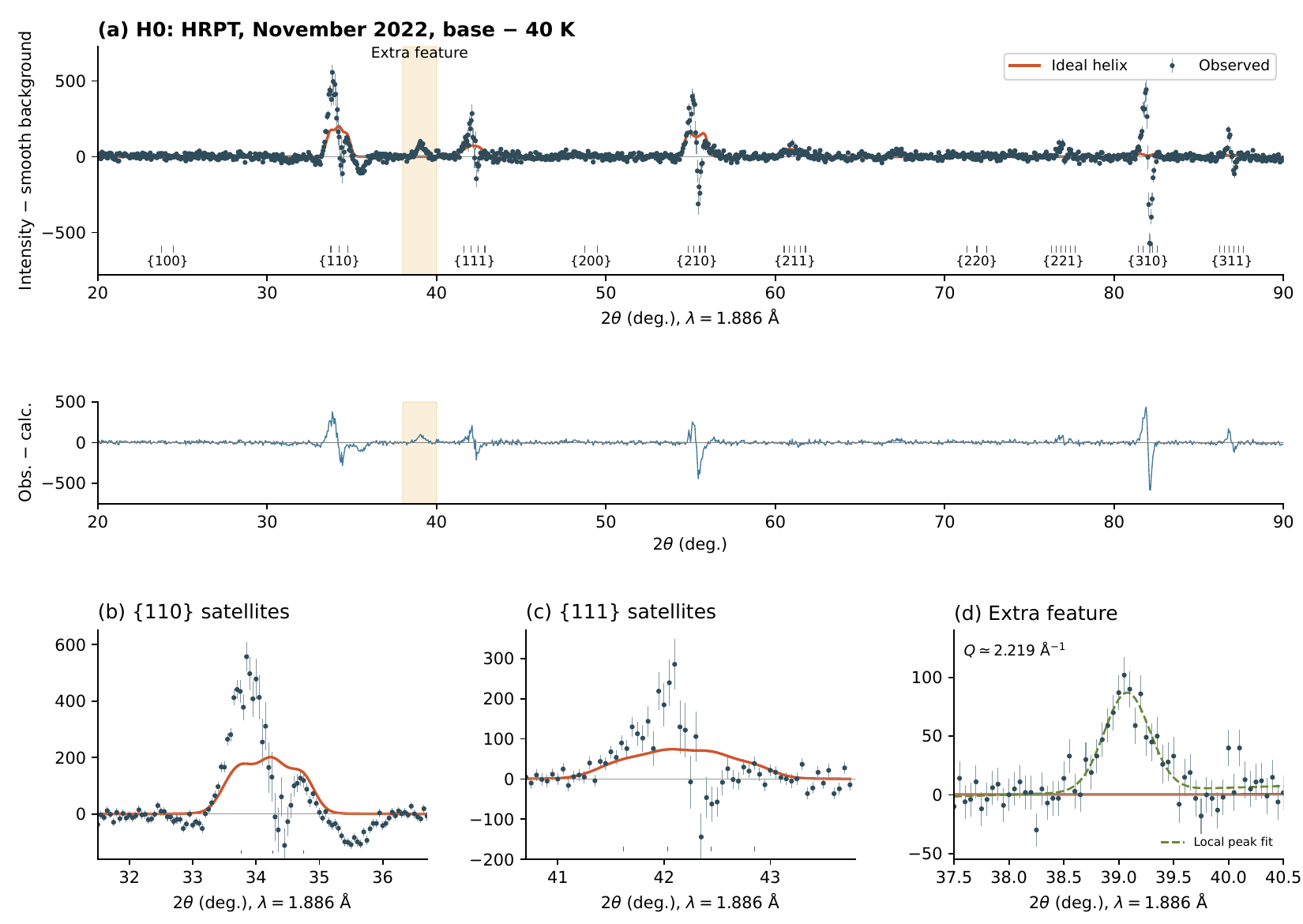}
\caption{Earlier HRPT dataset H0, measured in November 2022 at
$\lambda=1.886$~\AA: base-temperature minus 40~K data and an
ideal-helix fit. (a) Full fitted interval, with the observed-minus-calculated
difference below; (b,c) enlarged $\{110\}$ and $\{111\}$ regions;
(d) the extra feature near $Q=2.219$~\AA$^{-1}$.
Points carry propagated counting errors; red curves show the
MnSi ideal-helix calculation. The smooth fitted background is
removed from both. Ticks mark indexed magnetic satellites.
The shaded $38^\circ$--$40^\circ$ interval is excluded from the
MnSi fit. The dashed green curve in (d) is a separate local
Gaussian-plus-linear-background fit, shown only to locate the
extra feature. Residual nuclear contributions limit the magnetic
refinement of this historical dataset.}
\label{fig:SM_H0_ideal_feature}
\end{figure}

Using the fixed parameters from the 40~K nuclear refinement, the
ideal-helix fit varies its amplitude, $k$ and three background coefficients
over $20^\circ<2\theta<90^\circ$, excluding $38^\circ$--$40^\circ$.
It gives $k=0.01406(25)$ and $\chi^2=2.8055$ for 1358 data points
and five parameters (Fig.~\ref{fig:SM_H0_ideal_feature}). The mismatch
between the base-temperature and 40~K lattice parameters contributes
to incomplete cancellation of the nuclear Bragg peaks.

After angular correction, the extra feature at $39.08^\circ$ corresponds
to $Q\simeq2.219$~\AA$^{-1}$, or $51.3^\circ$ at $\lambda=2.453$~\AA,
and is not reproduced by the MnSi Bragg peak indexing. Residual nuclear
scattering limits H0 to a historical comparison and evidence for the
feature. The quoted uncertainty in $k$ is conditional on the fixed
nuclear-refinement parameters and excludes these subtraction effects.

\subsection{Later HRPT dataset H4 and the residual extra feature}
\label{sm:H4}
\begin{figure}[htbp]
\centering
\includegraphics[width=\textwidth]{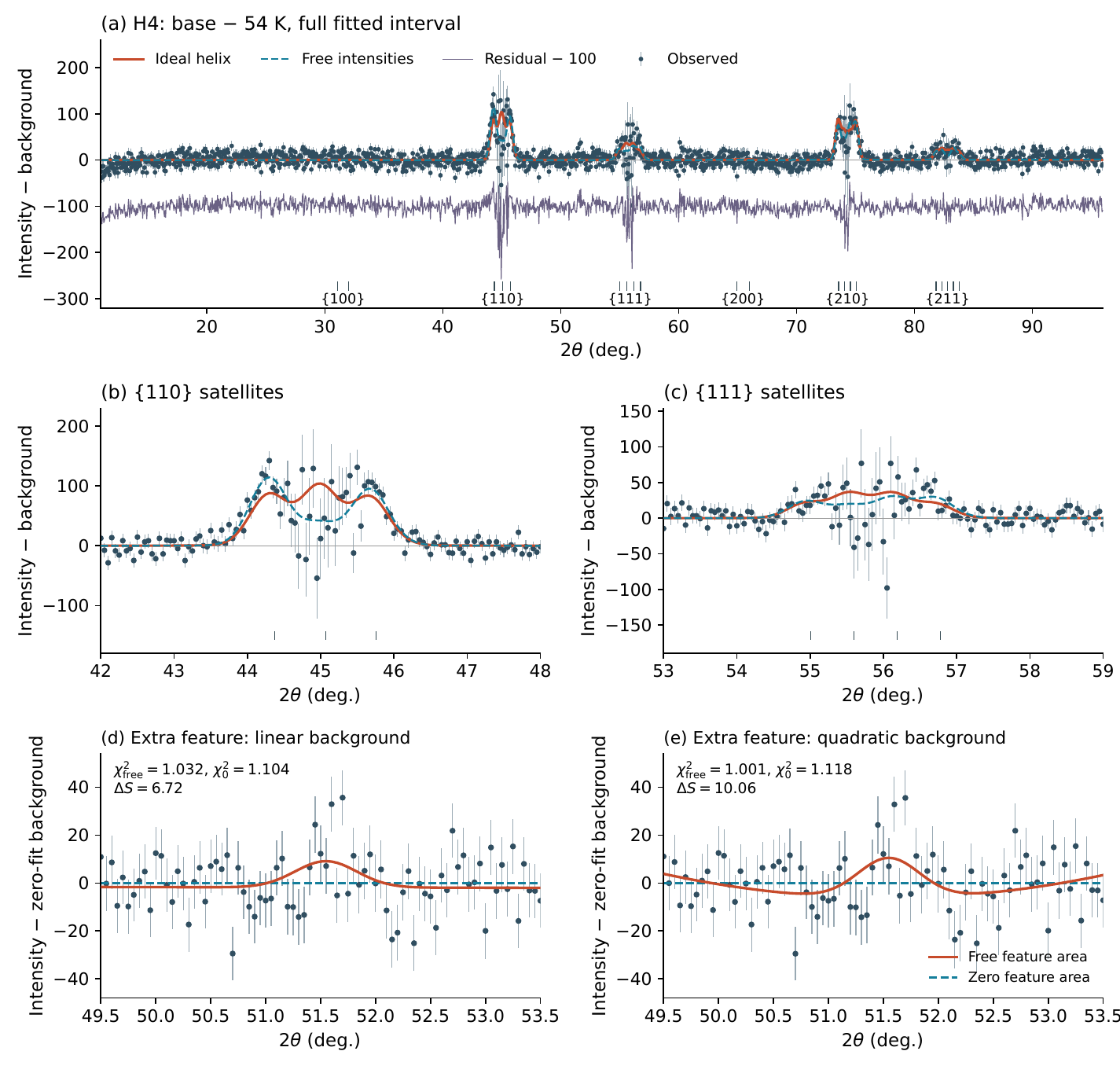}
\caption{Completed HRPT dataset H4: base-temperature minus 54~K
data points and fits. (a) Full interval; (b,c) the $\{110\}$ and
$\{111\}$ satellite regions. Red and dashed blue curves are the
ideal-helix and nonnegative free-intensity fits; the ideal-fit background
is subtracted from the measured intensities and both calculated curves.
The observed-minus-ideal residual is displaced downwards in (a).
Ticks mark indexed magnetic satellites, including the two $\{100\}$
positions whose intensities are fixed to zero.
(d,e) Local extra-feature fits with linear and quadratic backgrounds,
respectively, using the fixed H0-derived profile. The background from
the zero-area fit is subtracted from the measured intensities and both models
in each panel; the background is refitted when the area is freed.
Points retain the native $0.05^\circ$ spacing and propagated counting
errors. Intensities use the common monitor normalization of the H4
patterns. The local $\Delta S$ compares the zero-area and free-area fits.}
\label{fig:SM_H4_fits_feature}
\end{figure}

Dataset H4 was recorded a few weeks after the October DMC measurements,
after the powder container had been shaken. The same fitting procedures
use fixed parameters from its 54~K nuclear refinement over
$11^\circ<2\theta<96^\circ$, including the extra-feature region.
The ideal fit gives $k=0.01465(23)$ and
$m_{1k}=0.4354(61)\,\mu_{\rm B}$, or a constant moment
$0.533(8)\,\mu_{\rm B}$, consistent with $0.535(3)\,\mu_{\rm B}$ for H1--H3.
Figure~\ref{fig:SM_H4_fits_feature} and Table~\ref{tab:sm_profile_fits}
show the fits and statistics. One nonnegative group coefficient reaches
zero; the unrestricted complete $\{200\}$ area is consistent with zero.

The extra-feature test uses 81 bins over $49.5^\circ\leq2\theta\leq53.5^\circ$
and a bin-averaged Gaussian transferred from H0, with fixed centre
$51.549^\circ$ and FWHM $0.661^\circ$. Joint area--background fits give
$7.75\pm2.99$ (linear) and $11.79\pm3.72$ (quadratic), in the figure's
normalized intensity units multiplied by degrees. Errors include
background covariance. Setting the area to zero and refitting the
background raises $S$ from 80.46 to 87.18 (linear) and 77.11 to 87.17
(quadratic), with $\chi^2$ changing from 1.032 to 1.104 and 1.001 to 1.118,
respectively. The gains $\Delta S=6.72$ and 10.06 for one added parameter
correspond to conditional area-to-error ratios of 2.6 and 3.2.
Profiles transferred from October DMC give $\Delta S=4.97$--9.27.
These alternative fits favour a weak residual, with significance dependent
on profile and background assumptions; they are not independent detections.

Normalizing the earlier feature to MnSi nuclear areas predicts H4 areas
of 29--31 from H0 and 27--48 from October DMC, depending on the nuclear
reflection. An unchanged feature would therefore be readily detectable
under these transfer assumptions: H4 supports a strongly reduced feature,
not a loss caused solely by counting statistics. Coarse impurity grains
are compatible with this change.

\section{Single-arm symmetry, rhombohedral setting and mLD1 exclusion}
\label{sm:R3}
The single-arm mLD2 model has MSSG $R3.1'(0,0,g)ts$
(No.~146.1.22.2.m11.2). We use its standard $R$-centred hexagonal
setting, with zero origin shift and basis matrix
\begin{equation}
 P=\begin{pmatrix}0&1&-1\\1&-1&-1\\-1&0&-1\end{pmatrix},\qquad
 \mathbf r_c=P\mathbf r_h,\qquad
 \mathbf k_h=P^T\mathbf k_c=(0,0,-3k).
 \label{eq:sm_R3_basis}
\end{equation}
The subscripts $c,h$ denote parent cubic and hexagonal coordinates.
Thus $\mathbf a_h=\mathbf b_c-\mathbf c_c$,
$\mathbf b_h=\mathbf a_c-\mathbf b_c$ and
$\mathbf c_h=-\mathbf a_c-\mathbf b_c-\mathbf c_c$;
$a_h=\sqrt2a_c$, $c_h=\sqrt3a_c$ and $\gamma_h=120^\circ$.
The minus sign in $g=-3k$ follows from $\mathbf c_h$ pointing opposite
to cubic $[111]$. The parent Mn $4a$ orbit splits into $3a$ and $9b$,
corresponding to one and three Mn sites per parent primitive cell.

In coordinates $(x_h,y_h,z_h,t)$, the threefold generator is
$(-y_h,x_h-y_h,z_h,t+1/3;+1)$. The $R$-centring translations are
$(1/3,2/3,2/3;0)$ and $(2/3,1/3,1/3;0)$, and the antiunitary
internal translation is $(0,0,0;1/2;-1)$. 
The modulation on each representative site is
$\mathbf M_s(t)=\mathbf C_s\cos(2\pi t)+\mathbf S_s\sin(2\pi t)$.
The coefficients are components along the \emph{unit} hexagonal
crystallographic axes, which are not an orthogonal Cartesian basis.
On the axial $3a$ site, symmetry imposes
\begin{equation}
 C_z=S_z=0,\qquad S_x=(C_x-2C_y)/\sqrt3,\qquad
 S_y=(2C_x-C_y)/\sqrt3.
 \label{eq:sm_R3_constraints}
\end{equation}
This leaves two real coefficients. The general $9b$ representative
has six independent coefficients, for a total of eight, or seven
after choosing the unobservable common modulation phase.

For an explicit conversion to cubic Cartesian components, use
\begin{equation}
 \mathbf A_s=U\mathbf H_s,\quad
 \mathbf H_s=\mathbf C_s+i\mathbf S_s,\quad
 U=\begin{pmatrix}
 0&1/\sqrt2&-1/\sqrt3\\
 1/\sqrt2&-1/\sqrt2&-1/\sqrt3\\
 -1/\sqrt2&0&-1/\sqrt3
 \end{pmatrix}.
 \label{eq:sm_R3_vectors}
\end{equation}
With the general representative mapped to parent site
$(x+1/2,1/2-x,-x)$ and consistent site phases, the cubic coefficients are
\begin{equation}
 \begin{aligned}
 Z_0&=H_{1y}/\sqrt2,&
 Z_1&=H_{2y}/\sqrt2-H_{2z}/\sqrt3,\\
 Z_2&=(H_{2y}-H_{2x})/\sqrt2+H_{2z}/\sqrt3,&
 Z_3&=H_{2x}/\sqrt2+H_{2z}/\sqrt3.
 \end{aligned}
 \label{eq:sm_R3_mapping}
\end{equation}
Here $Z_j=m_j\exp(-2\pi i\alpha_j)$ uses the main-text real-space
phase convention. Setting $w=\exp(-2\pi i/3)$, the complex vectors
at its four Mn sites are
\begin{equation}
 \mathbf v_1=Z_0(1,w,w^2),\quad
 \mathbf v_2=(Z_1,-Z_2,-Z_3),\quad
 \mathbf v_3=w(-Z_3,Z_1,-Z_2),\quad
 \mathbf v_4=w^2(-Z_2,-Z_3,Z_1).
 \label{eq:sm_cubic_vectors}
\end{equation}
These are precisely the first-arm terms of the main-text general
modulation formula. The indices on $m_j,\alpha_j$ label coefficient
pairs, not propagation-vector arms: obtaining the single-arm model
retains all four Mn sites with $m_j,\alpha_j$ and only their first-arm terms.

\subsection{Numerical equivalence check}
The conversion was checked using matched FullProf refinements of the
earlier H2 dataset in the standard $R3$ MSSG setting and with an
equivalently constrained $P1$ model in cubic axes. Both fits used
equivalent lattice and propagation vectors, consistent profile and
background parameters, and matching reflection sets, with seven magnetic
variables in each setting. The conventional $R$-centred hexagonal cell
contains three primitive cells; with the atomic-occupancy and scale
conventions used in these PCR files, the corresponding scale conversion
is $s_h=s_c/9$. Here $s_c=85.95689$ and $s_h=9.55076556$.
This normalization conversion leaves the magnetic moments unchanged.
\begin{table}[htbp]
\caption{Matched single-arm refinements in two coordinate settings.
Both use the same 1699 data points between $11^\circ$ and $96^\circ$
and the earlier H2 nuclear-refinement parameters, distinct from those
used in Table~\ref{tab:sm_profile_fits}.}
\label{tab:sm_R3_equivalence}
\centering
\begin{ruledtabular}
\begin{tabular}{lrr}
Cubic component $k$ & $\chi^2$, constrained $P1$ (cubic axes) & $\chi^2$, standard $R3$\\
0.0146743 & 1.1693 & 1.1693\\
0.0141670 & 1.1833 & 1.1833\\
\end{tabular}
\end{ruledtabular}
\end{table}
For the same structure expressed in the two settings, the calculated
profiles agree to the printed precision of $0.01$ counts. The coordinate
equivalence is exact and does not depend on whether the refinement
has a unique global minimum.

\subsection{Why the pure mLD1 model is excluded}
At $\mathbf Q=\mathbf k_1$, the threefold rotation leaves the
scattering vector unchanged. For mLD1 its character is $+1$, so the
structure factor satisfies $C_3\mathbf F=\mathbf F$. The invariant
vector is parallel to $[111]$, giving
$\mathbf F_\perp=\widehat{\mathbf Q}\times\mathbf F
\times\widehat{\mathbf Q}=\mathbf 0$. The conjugate satellite and the rotated arms obey
the same argument. This contradicts the observed small-angle magnetic
satellites cited in the main text; it does not assert extinction of
all wide-angle mLD1 reflections.

As a separate numerical check, on the native $0.125^\circ$ DMC grid
over $20^\circ$--$88^\circ$, excluding $50^\circ$--$53^\circ$,
the seven-parameter mLD1 fit gives $\chi^2=23.7834$, compared with
$15.9856$ for the one-parameter mLD2 ideal helix at the same fixed
lattice, angular and profile parameters, scale and background.
Multiple starts and final FullProf refinement were used.
The DMC residual limitations in Sec.~\ref{sm:DMC_diagnostics}
remain; the small-angle symmetry constraint is independent of that
profile comparison.

\section{Two-satellite extinction and the ideal-helix limit}
\label{sm:ideal}

\subsection{Phase convention and simultaneous extinction}
With a positive scattering exponential, define the cell-origin Fourier
coefficient by
\begin{equation}
 \mathbf m_{ns}=\mathbf S_s e^{-2\pi i\mathbf k\cdot\mathbf n}
                  +\mathbf S_s^*e^{2\pi i\mathbf k\cdot\mathbf n},
 \qquad
 \mathbf S_s=\tfrac12\mathbf v_s e^{-2\pi i\mathbf k\cdot\mathbf r_s}.
 \label{eq:sm_phase_convention}
\end{equation}
Here $\mathbf n$ is an integer lattice translation and $\mathbf r_s$
the fractional basis position. The fixed irrep phases are included in $\mathbf v_s$ of
Eq.~\ref{eq:sm_cubic_vectors}. At $\mathbf Q=\mathbf G+\mathbf k$,
the positional phase in $\mathbf S_s$ cancels the corresponding part
of the scattering exponential, leaving
$\mathbf F=\tfrac12\sum_s\mathbf v_s e^{2\pi i\mathbf G\cdot\mathbf r_s}$.
The positional phase may be absorbed into the Fourier coefficients, but the coefficients and scattering expression must use the same phase convention. In the cubic FullProf PCR representation,
the same convention is implemented by \texttt{MagPh}, including the
positional and fixed symmetry phases. The additional $\alpha_j$ are
independent coefficient phases.

For $\mathbf G=(100)$, set $a=\exp(2\pi i x)$ and $b=a^*$.
Omitting only common factors (the Fourier factor $1/2$ and magnetic
form factor), the structure factor is
\begin{equation}
 \begin{aligned}
 F_x&=a(Z_0-Z_1)+b(w^2Z_2-wZ_3),\\
 F_y&=a(wZ_0+Z_2)+b(wZ_1+w^2Z_3),\\
 F_z&=a(w^2Z_0+Z_3)-b(wZ_2+w^2Z_1).
 \end{aligned}
 \label{eq:sm_extinction_F}
\end{equation}
At $\mathbf Q_+=(1+k,k,k)$, extinction of the perpendicular
component requires $(1+k)F_y-kF_x=(1+k)F_z-kF_x=0$.
For the other powder group use $\mathbf G=(-100)$, interchange
$a,b$, and replace $1+k$ by $-1+k$. The two conditions together
form a complex linear system of rank three at the physical Mn
coordinate and propagation vector. Its one-dimensional null space is
\begin{equation}
 Z_1=Z_0,\qquad Z_2=e^{i\pi/3}Z_0,\qquad
 Z_3=e^{-i\pi/3}Z_0.
 \label{eq:sm_extinction_solution}
\end{equation}
Thus all $m_j$ are equal and
$(\alpha_1-\alpha_0,\alpha_2-\alpha_0,\alpha_3-\alpha_0)
=(0,-1/6,+1/6)$ modulo a cycle. These additional phases work
together with the symmetry-imposed $\pm1/3$ phases; they do not
replace them. All four site vectors become $Z_0(1,w,w^2)$,
transverse to $[111]$ and circular. Their positional sum at $(100)$
is $a-a+b-b=0$, so the full structure factor vanishes independently
of $x$ and $k$. Imposing only one of the two satellite extinctions
leaves a larger solution space and does not select this ideal limit.
Distinct-arm contributions add as nonnegative intensities, so exact
extinction cannot be produced by cancellation between different arms.
Finite experimental sensitivity defines a bound about this ideal
limit, not a proof of exact zero intensity.

\subsection{Moments and their uncertainties}
For a single circular helix, the instantaneous magnitude is
$M_{1k}=\sqrt{3/2}\,m_{1k}$. The powder-equivalent equal-arm
construction has $m_{4k}=m_{1k}/2$. The inter-arm cross terms
average to zero in the incommensurate structure, hence
$\langle|\mathbf M_{4k}|^2\rangle=6m_{4k}^2=M_{1k}^2$.
For the four tetrahedral unit propagation directions $\mathbf n_j$,
$\sum_j\mathbf n_j\mathbf n_j^T=(4/3)\mathbf 1$. Projecting each
arm of length $h=\sqrt{3/2}\,m_{4k}$ onto the direction $\mathbf e$
of the total moment gives
\begin{equation}
 |\mathbf M_{4k}|\leq h\sum_{j=1}^4
 \sqrt{1-(\mathbf e\cdot\mathbf n_j)^2}
 \leq h\sqrt{32/3}=4m_{4k}.
\end{equation}
With $\alpha_0=0$ and zero additional arm phases, as in the
main-text illustration,
the continuous modulation reaches zero and this upper bound; these
are limiting values for dense incommensurate atomic sampling.
The common phase is not determined by the  satellite
intensities. Changing it can alter the attained maximum, while the
upper bound $4m_{4k}$ remains valid.

The abstract's moment errors use the independent native FullProf
amplitudes $m_{1k}=0.4410(56)$, $0.4426(59)$ and
$0.4335(36)\,\mu_{\rm B}$ for H1, H2 and H3, respectively,
with $k$, the nuclear-refinement parameters and nuclear scale fixed.
Their formal errors differ slightly from those of the amplitude--$k$
fits in the main-text comparison table.
For weights $w_d=1/\sigma_{m,d}^2$,
$\overline m=\sum_d w_dm_d/\sum_dw_d$ and
$\sigma_{\overline m}=(\sum_dw_d)^{-1/2}$ give
$\overline m_{1k}=0.4371(27)\,\mu_{\rm B}$.
Here \(d\) labels the independent datasets H1, H2 and H3.
Linear propagation then gives $M_{1k}=0.535(3)\,\mu_{\rm B}$ and
$M_{4k}^{\rm max}=2\overline m_{1k}=0.874(5)\,\mu_{\rm B}$.
These are conditional statistical errors, excluding nuclear-scale,
profile-calibration and subtraction uncertainties.

\section{Exploratory departures from the ideal helix}
\label{sm:departures}
For the earlier H2 dataset of Sec.~\ref{sm:R3}, the matched
ideal-helix and seven-variable refinements give $\chi^2=1.1785$
and 1.1693, respectively, at fixed $k=0.0146743$ on the same
1699 data points. The fixed cell, angular and profile parameters,
background, scale and reflection support are identical.

The seven-variable search fixes the common phase through $I_0=0$
and varies $\mathbf v=(R_0,R_1,R_2,R_3,I_1,I_2,I_3)^T$.
With the other parameters fixed, each magnetic structure factor is
linear in $\mathbf v$, so the calculated intensity at data point $i$
is a quadratic form:
\[
 I_i^{\rm calc}=b_i+\mathbf v^T A_i\mathbf v,\qquad
 S(\mathbf v)=\sum_i\left[
 \frac{b_i+\mathbf v^T A_i\mathbf v-I_i^{\rm obs}}{\sigma_i}
 \right]^2 .
\]
Here $b_i$ is the fixed background and $A_i$ is a symmetric $7\times7$
matrix incorporating the FullProf structure factors and profiles.
The matrices were reconstructed from 29 FullProf calculations:
the zero vector, seven unit coefficient vectors and their 21 pairwise
sums. This represents the quadratic intensity dependence without
linearizing it about a trial structure, subject to the precision of
the exported profiles. The symmetry-allowed $\{100\}$ intensities
remain part of this model; the ideal-helix extinctions are not imposed.

We minimized $S$ iteratively with the trust-region algorithm
\texttt{trf} in SciPy's \texttt{least\_squares}, using no bounds and
the analytic residual Jacobian
$J_{ij}=2(A_i\mathbf v)_j/\sigma_i$.
The trust-region radius adapts to the agreement between the predicted
and actual reduction of $S$. This is nonlinear fitting of magnetic
coefficients, in contrast to the linear fit of independent group
intensities in Sec.~\ref{sm:weighted}.
All 32 random starts converged: 18 reached one minimum and 14 another.
Direct FullProf refinements of these solutions gave $\chi^2=1.1693$
and 1.1701, respectively. The lower-residual solution is shown in
Fig.~4 of the main text and Table~\ref{tab:sm_general_coefficients}.

As an independent check, FullProf was also started directly from the
same 32 random sets. With relaxation factor 0.2, 30 runs converged
within 400 cycles; the two divergent runs converged when restarted
with factor 0.05. These runs recovered the same two minima.
The two solutions have different coefficients but very similar powder
profiles; reproducible convergence does not establish a unique nonideal
magnetic structure or prove that the global minimum has been found.

Table~\ref{tab:sm_general_coefficients} also gives an earlier
simulated-annealing (SA) solution and its six-variable least-squares
continuation, distinct from the 32-start test. All coefficients use
the single-arm normalization of the main text.
The complex amplitudes are $Z_j=R_j+iI_j=m_j\exp(-2\pi i\alpha_j)$,
so $m_j=(R_j^2+I_j^2)^{1/2}$ and
$\phi_j=360^\circ\alpha_j=-\arg Z_j$, with the argument expressed
in degrees. The $R_j,I_j$ and $\phi_j$ columns retain each solution's
original common-phase convention.
The final column gives $\Delta\phi_j=\phi_j-\phi_0=-\arg(Z_j/Z_0)$,
reduced to $[-180^\circ,180^\circ)$, corresponding to a phase convention
with $\phi_0=0$. Thus $\Delta\phi_0=0$ by definition. The measured
intensities constrain these relative coefficient phases, not the common
phase. The ideal-helix values are
$(\Delta\phi_1,\Delta\phi_2,\Delta\phi_3)=(0^\circ,-60^\circ,60^\circ)$.

The six-variable fit converged after eight cycles with $R_0,R_1$ fixed.
Its additional constraint and historical low-angle profile treatment
distinguish it from the matched seven-variable test, so their
$\chi^2$ values are not directly comparable.
SA minimized $R_p$, whereas least squares minimized the weighted
squared residual.

For both least-squares solutions, the $R_j,I_j$ errors are taken from
FullProf; approximate errors on $m_j$, $\phi_j$ and $\Delta\phi_j$ are
propagated with their full covariance. In particular,
$\operatorname{var}(\Delta\phi_j)=\operatorname{var}(\phi_j)
+\operatorname{var}(\phi_0)-2\operatorname{cov}(\phi_j,\phi_0)$.
These formal uncertainties are conditional on the fixed lattice,
angular and profile parameters, nuclear scale, background,
propagation vector and imposed coefficient constraints.
\begin{table}[htbp]
\caption{Exploratory H2 solutions in the main-text single-arm
normalization. $R_j,I_j$ and $m_j$ are coefficient amplitudes in
$\mu_{\rm B}$. Phases are in degrees; $\Delta\phi_j=\phi_j-\phi_0$
gives the equivalent phases with $\phi_0=0$, while $R_j,I_j$ and $\phi_j$
retain their original convention. Parentheses and $\pm$ give formal
one-standard-deviation errors, including covariance for the phase
differences. ``Fixed'' denotes an imposed fit constraint;
$\Delta\phi_0=0$ is the reference by definition. SA values are central
estimates without least-squares error bars.}
\label{tab:sm_general_coefficients}
\centering
\begin{ruledtabular}
\begin{tabular}{crrrrr}
$j$ & $R_j$ & $I_j$ & $m_j$ & $\phi_j$ ($^\circ$) & $\Delta\phi_j$ ($^\circ$)\\
\multicolumn{6}{l}{Matched seven-variable solution ($\chi^2=1.1693$)}\\
0 & $0.515(32)$ & $0$ (fixed) & $0.515(32)$ & $0$ (fixed) & $0$\\
1 & $0.361(34)$ & $0.066(89)$ & $0.367(38)$ & $-10.4\pm13.6$ & $-10.4\pm13.6$\\
2 & $0.216(90)$ & $0.502(43)$ & $0.546(54)$ & $-66.7\pm8.8$ & $-66.7\pm8.8$\\
3 & $0.236(68)$ & $-0.295(54)$ & $0.377(58)$ & $51.3\pm9.7$ & $51.3\pm9.7$\\
\multicolumn{6}{l}{Historical SA starting solution ($R_p\simeq0.886\%$)}\\
0 & $-0.06086$ & $-0.50252$ & $0.50619$ & $96.91$ & $0$\\
1 & $-0.06922$ & $-0.37813$ & $0.38442$ & $100.37$ & $3.47$\\
2 & $0.24572$ & $-0.27361$ & $0.36775$ & $48.07$ & $-48.83$\\
3 & $-0.46477$ & $-0.28440$ & $0.54488$ & $148.54$ & $51.63$\\
\multicolumn{6}{l}{Historical six-variable LSQ solution ($\chi^2=1.2712$)}\\
0 & $-0.06116$ (fixed) & $-0.503(34)$ & $0.507(34)$ & $96.93\pm0.47$ & $0$\\
1 & $-0.06857$ (fixed) & $-0.341(43)$ & $0.348(43)$ & $101.4\pm1.4$ & $4.4\pm1.6$\\
2 & $0.277(52)$ & $-0.195(58)$ & $0.339(60)$ & $35.2\pm8.4$ & $-61.8\pm8.5$\\
3 & $-0.506(43)$ & $-0.153(73)$ & $0.529(42)$ & $163.2\pm7.9$ & $66.2\pm7.8$\\
\end{tabular}
\end{ruledtabular}
\end{table}

The historical DMC annealing/least-squares comparison used a
$0.05^\circ$ rebinned pattern (this was because of the bug, which prevented natural step in SA minimization) over $20^\circ$--$88^\circ$, excluding
$50^\circ$--$53^\circ$. With the same nonmagnetic settings, the
ideal one-amplitude fit gives $\chi^2=14.8330$, and the
seven-variable continuation gives 11.8184. These are descriptive
statistics on that rebinned representation; they must not be compared
numerically with the native-grid value 15.9856 as if the data points
were identical. Rebinning does not create independent measurements.
The general fit also predicts nonzero intensities for both $\{100\}$
groups, unlike the ideal limit. Together with the native-grid
diagnostics in Sec.~\ref{sm:DMC_diagnostics}, these checks prevent
interpreting a lower global residual alone as a determination of
departures from the ideal helix.

\section{DMC temperature-dependent integrated areas}
\label{sm:temperature}

\begin{figure}[h]
\centering
\includegraphics[width=\textwidth]{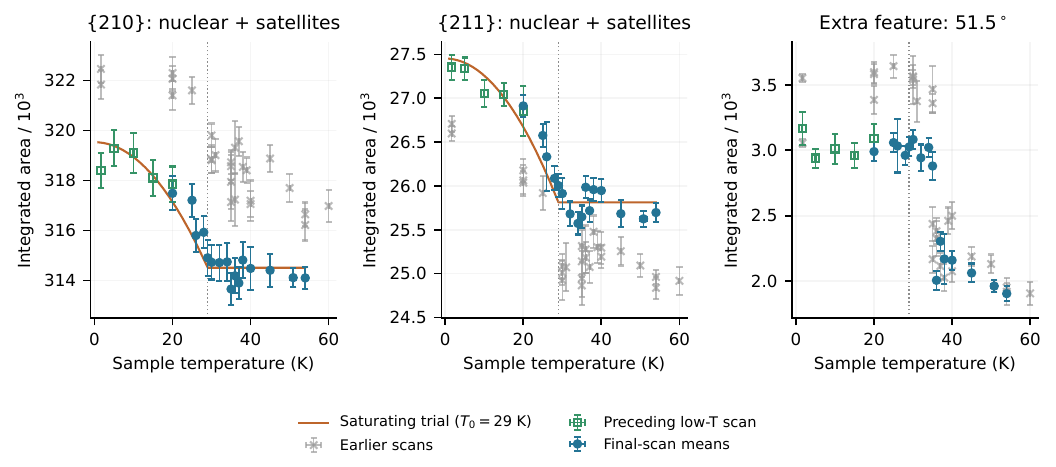}
\caption{Extended DMC temperature coverage on the original common
monitor scale. Blue points are the final-scan means; open green squares
are the preceding low-temperature sequence. Grey crosses show earlier scan histories for context
and are excluded from the orange saturating trial curves, which use
Eq.~\ref{eq:sm_Tsaturating} with $T_0=29$~K fixed.  Vertical bars are effective area uncertainties.}
\label{fig:sm_Thistories}
\vspace{1.5em}
\includegraphics[width=0.94\textwidth]{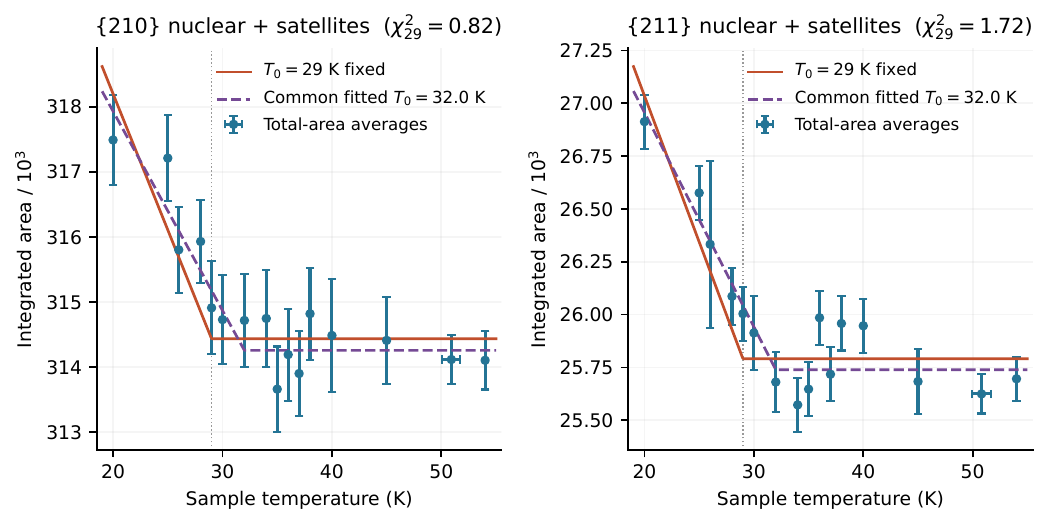}
\caption{Constant-plus-linear temperature trial for the total
$\{210\}$ and $\{211\}$ areas. Points are the subset of 16 final-scan
means, with effective errors. The quoted $\chi^2$ values refer to the fixed-29~K
fits.}
\label{fig:sm_Tlinear}
\end{figure}

The $\{210\}$ and $\{211\}$ areas in the main-text temperature plot
contain nuclear scattering and unresolved magnetic satellites together.
They are estimated from a flexible nonnegative Gaussian envelope plus
a jointly fitted local linear background.  Its area is integrated over the windows in
Table~\ref{tab:sm_Twindows}. 
\begin{table}[htbp]
\caption{Signal and background intervals in $2\theta$ for the main-text
temperature figure, in degrees. The $51.5^\circ$ band is a direct integral;
the other two entries use fitted envelope areas.}
\label{tab:sm_Twindows}
\centering
\begin{ruledtabular}
\begin{tabular}{lrrr}
Region & Signal & Lower background & Upper background\\
$\{210\}$ & 72--76 & 70.5--71.5 & 76.5--77.5\\
$\{211\}$ & 81--85 & 79.5--80.5 & 85.5--87\\
$51.5^\circ$ band & 50.5--52.15 & 49.5--50.125 & 52.5--53.4\\
\end{tabular}
\end{ruledtabular}
\end{table}
Areas retain the common monitor normalization and angular integration
step. No 54~K pattern is subtracted, so these area estimates do
not share a subtracted reference. The fitted-area covariance includes
the joint background parameters.

Figure~\ref{fig:sm_Thistories} compares the separate scan histories.
At comparable temperatures, the earlier measurements (grey crosses)
generally lie above the final-scan data for $\{210\}$ and below them
for $\{211\}$. These opposite shifts are consistent with possible
measurement instabilities and cannot be accounted for by a single
common normalization factor. The earlier scans are therefore retained
as a reproducibility check and are not pooled with the final scan
for the temperature fits below.

The orange curves in main-text Fig.~8(a,b) and
Fig.~\ref{fig:sm_Thistories} fit these five data points and the
16 final-scan means together without any relative rescaling, using
\begin{equation}
 I(T)=C+D\max[1-(T/T_0)^2,0],\qquad T_0=29~{\rm K}.
 \label{eq:sm_Tsaturating}
\end{equation}
This empirical function permits a nearly constant low-temperature
intensity. It gives $\chi^2=0.89$ and 1.83 for $\{210\}$ and
$\{211\}$ ($N=21$, $n_{\rm par}=2$), compared with 1.70 and 3.31
for a linear term extended to the same low temperatures.

At the repeated 20~K target, the earlier grey $\{210\}$ areas in
Fig.~\ref{fig:sm_Thistories} differ from the final scan by approximately
1--1.5\%, comparable to the excess being estimated. Averaging all
periods would therefore confound measurement history with temperature.

To test a simple temperature description, the total areas were fitted to
$ I(T)=C+A\max(T_0-T,0)$. The constant $C$ approximates the nuclear contribution, while the
linear term describes a possible magnetic excess below the fixed
temperature. This assumption is tested on the total integrated areas;
it does not independently separate satellites from nuclear scattering.
The fits use the 16 final-scan means over 20--54~K and their effective
errors, with two parameters for each region (Fig.~\ref{fig:sm_Tlinear}).

Between 25 and 30~K the $51.5^\circ$ band areas are 3057.9 and
3082.0, giving a contrast $24.1\pm102.4$ on the common intensity
scale. Normalizing the contrast and its error to the 25~K area gives
approximately $(0.8\pm3.3)\%$: no significant change is resolved.

\section{Common continuous fields and lower-arm superspace groups}
\label{sm:continuous_lower_arms}

\subsection{Inherited coefficient relations and the common-field condition}
\label{sm:common_field_condition}

Write the complex mode amplitudes of main-text Eq.~(3) as
$p_j=m_j\exp(2\pi i\alpha_j)$, $j=0,\ldots,3$, and let
$t_\nu=x_{\nu+3}$ denote the internal coordinates in cycles.
The contribution of the first arm at Mn site $s$ is
\begin{equation}
 \mathbf M_s^{[1]}(t_1)=\operatorname{Re}
 \left[\mathbf h_s e^{2\pi i t_1}\right],\qquad
 \mathbf h_s=(h_{sx},h_{sy},h_{sz}).
 \label{eq:sm_first_arm_coefficients}
\end{equation}
Here $p_j=Z_j^*$, where $Z_j$ is defined in Sec.~\ref{sm:R3};
the conjugation reflects the opposite sign of the Fourier exponential
and does not change the real modulation.

For the two-arm $\{1,2\}$, three-arm $\{1,2,3\}$ and fixed-section
four-arm MSSGs in Table~\ref{tab:sm_lower_groups}, no nonidentity
spatial point operation leaves $\mathbf k_1$ fixed. Its complex
three-vector may therefore be specified independently on each of the
four Mn sites, with the other arms generated by the respective group
operations. This gives $4\times3=12$ complex, or 24 real,
first-harmonic coefficients. The symmetry-related $\{2,3,4\}$ domain
has the same count after relabelling.

Equation~(3) retains additional relations. With
$\omega=\exp(2\pi i/3)$ and the cyclic component permutation
$\mathcal C(v_x,v_y,v_z)=(v_z,v_x,v_y)$, they are
\begin{equation}
 \begin{aligned}
 \mathbf h_1&=p_0(1,\omega,\omega^{-1}),&
 \mathbf h_2&=(p_1,-p_2,-p_3),\\
 \mathbf h_3&=\omega\mathcal C\mathbf h_2,&
 \mathbf h_4&=\omega^{-1}\mathcal C^2\mathbf h_2.
 \end{aligned}
 \label{eq:sm_inherited_coefficients}
\end{equation}
The component relations at Mn1 impose two complex constraints;
those fixing Mn3 and Mn4 from Mn2 impose another six. Hence
\begin{equation}
 12-(2+6)=4\ \text{complex coefficients}
 \quad\Longleftrightarrow\quad 8\ \text{real coefficients}.
 \label{eq:sm_coefficient_count}
\end{equation}
The remaining parameters are precisely $p_0,p_1,p_2,p_3$.
These counts precede any choice of the common phase.

A common field on all four Mn sites further requires
$\mathbf h_1=\mathbf h_2=\mathbf h_3=\mathbf h_4$.
Equating the first two vectors in Eq.~\ref{eq:sm_inherited_coefficients}
gives three independent complex constraints,
\begin{equation}
 (p_0,p_1,p_2,p_3)=p_0(1,1,e^{-i\pi/3},e^{i\pi/3}).
 \label{eq:sm_common_condition}
\end{equation}
The remaining two vectors then agree automatically. Substitution
verifies the same equality for every arm, covering all 48
site/component/arm coefficients. Thus one complex amplitude remains:
its magnitude and common phase. These are exactly the ideal-helix
relations of main-text Eq.~(4), and they yield the common field of
Eq.~(5). Generic MSSG-allowed coefficients need not satisfy them:
differences between Mn basis sites can persist as $k\to0$.
The condition concerns a slowly varying field of the retained
harmonics, rather than an arbitrary interpolation between discrete moments.

\subsection{One-, two- and three-arm ideal fields}

Using the same common phase $\alpha_0$ as in the main text, define
\begin{equation}
 \theta_\nu=2\pi(t_\nu+\alpha_0),\qquad
 c_\nu=\cos\theta_\nu,\qquad
 c_{\nu\pm}=\cos(\theta_\nu\pm2\pi/3).
 \label{eq:sm_arm_cosines}
\end{equation}
Thus $c_\nu=A_{\nu+3}$ and $c_{\nu\pm}=A_{\pm(\nu+3)}$ in the
notation of main-text Eq.~(5). Both $t_\nu$ and $\alpha_0$ are in
cycles, whereas $\theta_\nu$ is in radians.
Choosing the active arms to be $\{1\}$, $\{1,2\}$ and
$\{2,3,4\}$ gives, respectively,
\begin{align}
 \mathbf M^{(1k)}&=m_{1k}
 \begin{pmatrix}c_1\\c_{1+}\\c_{1-}\end{pmatrix},
 \label{eq:sm_field1}\\
 \mathbf M^{(2k)}&=m_{2k}
 \begin{pmatrix}c_1+c_2\\c_{1+}-c_{2+}\\c_{1-}-c_{2-}\end{pmatrix},
 \label{eq:sm_field2}\\
 \mathbf M^{(3k)}&=m_{3k}
 \begin{pmatrix}c_2-c_3-c_4\\-c_{2+}+c_{3+}-c_{4+}\\-c_{2-}-c_{3-}+c_{4-}\end{pmatrix}.
 \label{eq:sm_field3}
\end{align}
The three-arm field in Eq.~\ref{eq:sm_field3} omits $\mathbf k_1$,
as in the illustration below. The choice $\{1,2,3\}$ used for the
ISODISTORT comparison in Sec.~\ref{sm:restricted_groups} is a symmetry-related domain,
with $\mathbf k_4$ omitted instead. Omitting an arm removes all its
terms on every Mn site; it does not mean setting one of the mode
amplitudes $m_j$ in Eq.~(3) to zero.

These functions are shared by all four Mn sites. Let $\mathbf R$ denote
physical position in cubic Cartesian axes and
$\mathbf X=(X,Y,Z)=\mathbf R/a$ its fractional crystallographic coordinates.
At the discrete Mn sites, $\mathbf X=\mathbf n+\mathbf r_j$, where
$\mathbf n$ is an integer translation and $\mathbf r_j$ a fractional
basis position. For zero additional arm-phase offsets, the continuous section is
\begin{equation}
 \begin{gathered}
 \tilde{\mathbf r}=(\tilde x,\tilde y,\tilde z)=2\pi k\,\mathbf X,
 \qquad t_\nu=\mathbf k_\nu\cdot\mathbf X,\\
 2\pi t_\nu=(\mathbf k_\nu/k)\cdot\tilde{\mathbf r},
 \qquad \nu=1,\ldots,4.
 \end{gathered}
\end{equation}
Here $k>0$ is the common magnitude of the cubic components of the
propagation vectors in reciprocal-lattice units. The $t_\nu$ are in cycles, whereas
the tilded coordinates are in radians. For example,
$2\pi t_1=\tilde x+\tilde y+\tilde z$.
For two arms, defining the radian variables $\tilde u=\tilde x+2\pi\alpha_0$
and $\tilde v=\tilde y+\tilde z$ gives the useful product form
\begin{equation}
 \mathbf M^{(2k)}=2m_{2k}
 \begin{pmatrix}
 \cos\tilde u\cos\tilde v\\
 -\sin(\tilde u+2\pi/3)\sin\tilde v\\
 -\sin(\tilde u-2\pi/3)\sin\tilde v
 \end{pmatrix}.
 \label{eq:sm_field2_product}
\end{equation}
It is independent of $\tilde y-\tilde z$: a view along $[0\bar11]$ therefore
contains both propagation directions in its screen plane, and the
continuum field does not vary with depth. For three arms,
$\mathbf M^{(3k)}/m_{3k}$ is equivalently the dimensionless four-arm
field of the main text minus $\mathbf M^{(1k)}/m_{1k}$ from
Eq.~\ref{eq:sm_field1}.

Each arm is a circular transverse helix of constant magnitude
$\sqrt{3/2}\,m_{Nk}$. For $N$ active arms the spatial mean is
\begin{equation}
 \langle|\mathbf M^{(Nk)}|^2\rangle=\tfrac32 N m_{Nk}^2.
\end{equation}
Consequently $m_{Nk}=m_{1k}/\sqrt N$ permits visual comparisons with
the same mean-square moment without normalizing the local arrows.
The one-, two- and three-arm vectors are linearly independent, so
their phases can be shifted by a real-space translation. The fourth
arm introduces the translation-invariant sum of phases discussed in
Sec.~V of the main text. These lower-arm examples are model
constructions, not additional refinements of the diffraction data.

\subsection{Identification of the restricted superspace groups}
\label{sm:restricted_groups}

We retained the operations of main-text Table~I that map the selected
arms into themselves, restricted their internal-coordinate matrices
to those arms, and included time reversal accompanied by a half-period
shift in every retained internal coordinate. The operator sets close
exactly modulo lattice translations. Direct substitution also verifies
the covariance of Eq.~(3) for arbitrary complex mode amplitudes.
Their standard groups were identified with FINDSSG in the ISOTROPY
suite; the results are listed in Table~\ref{tab:sm_lower_groups}.
The extended cubic parent uses the non-tabulated symbol
$P2_1 3\,1'(a,a,a)\,0{-t}s\allowbreak
(a,-a,-a)\,0{-t}s\allowbreak
(-a,a,-a)\,0{-t}s\allowbreak
(-a,-a,a)\,0{-t}s$ introduced in Sec.~IV B. Here $a$ is a
modulation-vector component, $t=1/3$ and $s=1/2$; the definitive
operations are those of main-text Table~I. This extended symbol
describes the family of phase sections and must be distinguished
from the tabulated fixed-phase-sum four-arm group in the last row
of Table~\ref{tab:sm_lower_groups}.
For three arms, an independent ISODISTORT mCIF export also gives
exactly the same affine operations after the setting transformation
below, including all 18 operations in the conventional $R$-centred cell.
\begin{table}[htbp]
\caption{Tabulated MSSGs of the restricted-arm families. The final row
is a four-arm model with three independent internal coordinates, not
the three-active-arm model in the preceding row. Standard-setting
parameters $a,b,g$ specify modulation-vector components. The three-arm choices $\{1,2,3\}$ and
$\{2,3,4\}$ give the same MSSG type. The two-arm entry uses the
standard unique-$c$ setting. For the first-harmonic Mn modulation,
MSSG symmetry alone permits eight real coefficients in the first row
and 24 in each remaining row; Eq.~(3) retains eight in every case,
before the ideal-helix constraints are imposed.}
\label{tab:sm_lower_groups}
\centering
\begin{ruledtabular}
\begin{tabular}{lll}
Active arms & Standard magnetic superspace symbol & Identifier\\
$1$ & $R3.1'(0,0,g)ts$ & 146.1.22.2.m11.2\\
$1,2$ & $P2_1.1'(a,b,g)0s(-a,-b,g)0s$ & 4.2.16.7.m8.2\\
$1,2,3$ & $R3.1'(a,b,g)0s(-a-b,a,g)0s(b,-a-b,g)0s$ & 146.3.185.3.m11.2\\
$1,2,3,4$ & $P2_12_12_1.1'(a,b,g)000s(a,-b,-g)000s(-a,b,-g)000s$ & 19.3.146.7.m26.2
\end{tabular}
\end{ruledtabular}
\end{table}

In the following setting transformations, $\mathbf r=(r_x,r_y,r_z)^T$
denotes parent cubic fractional coordinates. The primed vectors
$\mathbf r'_{2k}$ and $\mathbf r'_{3k}$ describe the same physical
position in the respective standard settings. Both transformations
start from the parent cubic coordinates; they are not successive steps.
For two arms the nontrivial spatial generator in cubic coordinates is
\begin{equation}
 (r_x,r_y,r_z;t_1,t_2)\mapsto
 (r_x+\tfrac12,-r_y+\tfrac12,-r_z;t_2,t_1).
\end{equation}
The transformation $\mathbf r'_{2k}=(r_y+3/4,r_z,r_x)^T$ converts it to
the standard $2_1$ screw along the new $c'$ axis; the modulation vectors
in this setting become $(k,k,k)$ and $(-k,-k,k)$. The internal coordinates
are unchanged, $t'_\nu=t_\nu$ for $\nu=1,2$.

For the three-arm choice $\{1,2,3\}$, the transformation to the
$R$-centred hexagonal setting is
\[
 P_{3k}=\begin{pmatrix}
 0&-1&1\\
 -1&1&1\\
 -1&0&-1
 \end{pmatrix},\qquad
 \mathbf p_{3k}=\begin{pmatrix}0\\1/2\\0\end{pmatrix},\qquad
 \mathbf r=P_{3k}\mathbf r'_{3k}+\mathbf p_{3k}.
\]
The columns of $P_{3k}$ give the new basis vectors in parent cubic
coordinates, and $\mathbf p_{3k}$ is the origin shift in the same
fractional coordinates. Thus
$\mathbf r'_{3k}=P_{3k}^{-1}(\mathbf r-\mathbf p_{3k})$.
For the internal coordinates, take
$(t'_1,t'_2,t'_3)=(t_3,t_1+2/3,t_2+1/3)$.
Writing $\mathbf r'_{3k}=(r'_x,r'_y,r'_z)^T$, one retained threefold
generator then has the standard form
\begin{equation}
 (r'_x,r'_y,r'_z;t'_1,t'_2,t'_3)\mapsto
 (-r'_x+r'_y,-r'_x,r'_z;t'_2,t'_3,t'_1).
\end{equation}
The modulation vectors become $(0,2k,k)$, $(-2k,0,k)$ and
$(2k,-2k,k)$, in that order. Together with the $R$-centering
translations and time reversal combined with the internal shift
$(1/2,1/2,1/2)$, these operations reproduce the ISODISTORT export.
The internal-origin offsets depend on the starting phase convention;
they do not change the MSSG type. The single-arm transformation
was given in Sec.~\ref{sm:R3}.

The one-arm restriction of Eq.~(3) spans all eight real first-harmonic
Mn coefficients allowed by its group. For the other groups,
Eq.~\ref{eq:sm_inherited_coefficients} selects an eight-coefficient
subfamily from the 24 allowed by MSSG symmetry, as derived in
Sec.~\ref{sm:common_field_condition}. The three-arm and four-arm
ISODISTORT exports independently confirm the unrestricted count:
each lists eight real magnetic modes from each of mLD2LE2, mLD1LE1
and mLD3LE3. Specifying the MSSG alone therefore does not impose
the selected mLD2 family.

The inherited relations originate from the four-arm threefold
operation about $[111]$, which fixes Mn1 and $\mathbf k_1$ while
cycling Mn2--Mn4 and the other three arms, with the accompanying
internal translation. This operation no longer preserves the
two- or three-arm sets; for the four-arm orthorhombic description,
it changes the fixed phase sum, as shown below. Deleting arms or
eliminating the dependent coordinate in Eq.~(3) nevertheless retains
its coefficient relations. These are additional parameter constraints,
not additional symmetries of the reduced structure. The common-field
condition of Eq.~\ref{eq:sm_common_condition} is a further restriction.
Agreement with the tabulated MSSG establishes compatibility, not
the generality of the restricted model.

\subsubsection{Four-arm reduction to three independent internal coordinates}

The four-arm model in the last row of Table~\ref{tab:sm_lower_groups}
requires a different reduction from deleting an arm. Write
$t_\nu=x_{\nu+3}$, with all internal coordinates expressed in cycles.
At a position with cubic fractional coordinates $\mathbf X$, a physical
section is specified by constant offsets $\eta_\nu$:
\[
 t_\nu(\mathbf X)=\mathbf k_\nu\cdot\mathbf X+\eta_\nu,
 \qquad
 \Sigma=\sum_{\nu=1}^{4}t_\nu(\mathbf X)
       =\sum_{\nu=1}^{4}\eta_\nu\pmod1,
\]
because $\sum_\nu\mathbf k_\nu=0$. Thus $\Sigma$ is constant
throughout space for a given configuration; this does not prescribe
the same value for every configuration. Eliminating the dependent
coordinate means imposing
\[
 x_7=\Sigma-x_4-x_5-x_6\pmod1.
\]
A real-space translation changes the four phases by
$\mathbf k_\nu\cdot\Delta\mathbf X$, whose sum is zero, and therefore
cannot change $\Sigma$. The offsets $\eta_\nu$ specify the section;
they are distinct from the coefficient phases $\alpha_j$ in Eq.~(3).
For the ideal-helix relations, a common coefficient phase $\alpha_0$
adds $4\alpha_0$ to the total four-arm phase sum. Fixing the coordinate
sum therefore does not remove this coefficient-phase freedom.
For the zero-sum section used in the ISODISTORT comparison,
\begin{equation}
 \Sigma=x_4+x_5+x_6+x_7=0\pmod1,
 \qquad x_7=-x_4-x_5-x_6\pmod1.
 \label{eq:sm_fixed_section}
\end{equation}
Eliminating $x_7$ means making this substitution in every modulation
term, including the fourth-arm contribution. To distinguish this from
deleting an arm, denote the contribution of arm $\nu$ at Mn site $j$
by $\mathbf F_{j\nu}(u)$. On the section of
Eq.~\ref{eq:sm_fixed_section}, the two constructions are
\[
\begin{aligned}
 \mathbf M_j^{\mathrm{four\ arms}}(x_4,x_5,x_6)
 &=\sum_{\nu=1}^{3}\mathbf F_{j\nu}(x_{\nu+3})
   +\mathbf F_{j4}(-x_4-x_5-x_6),\\
 \mathbf M_j^{\mathrm{three\ arms}}(x_4,x_5,x_6)
 &=\sum_{\nu=1}^{3}\mathbf F_{j\nu}(x_{\nu+3}).
\end{aligned}
\]
Both use three independent internal coordinates, but only the second
deletes all $x_7$-dependent terms from Eq.~(3). In the first, the fourth
wave remains at $\mathbf k_4=-\mathbf k_1-\mathbf k_2-\mathbf k_3$,
with internal Fourier indices $(-1,-1,-1)$. For the restricted models
considered here, the four-arm fixed-section reduction gives the
$P2_12_12_1$-based MSSG, whereas deleting arm 4 gives the three-arm
$R3$-based MSSG. These are the distinct final two rows of
Table~\ref{tab:sm_lower_groups}; the notation $3+3$D alone does not
distinguish them.

In main-text Table~I, every internal-coordinate transformation is a
permutation followed by a translation $\boldsymbol\tau$. Consequently,
\begin{equation}
 \Sigma'=\Sigma+\sum_{\nu=1}^4\tau_\nu\pmod1,
 \qquad
 \sum_{\nu=1}^4\tau_\nu=0\pmod1
 \quad\text{for a section-preserving operation}.
 \label{eq:sm_section_test}
\end{equation}
One must first retain the operations satisfying this condition and
then substitute Eq.~\ref{eq:sm_fixed_section}; simply dropping the
seventh coordinate from every operation would not give the symmetry
of the fixed section. Operations 1, 2, 11 and 12 pass this test.
Their reduced forms are listed in Table~\ref{tab:sm_four_arm_reduction},
alongside the corresponding operation numbers in the ISODISTORT export.

\begin{table}[htbp]
\caption{Explicit reduction of the four-arm operations on
$x_4+x_5+x_6+x_7=0$ modulo one. The six-coordinate transformations
coincide exactly with the algebraic operation list of the saved
ISODISTORT model, in its exported setting. The last entry is the
time-reversal sign.}
\label{tab:sm_four_arm_reduction}
\centering
\begin{ruledtabular}
\begin{tabular}{ccl}
Main-text Table~I & ISODISTORT & Reduced operation\\
1 & 1 & $(x_1,x_2,x_3,x_4,x_5,x_6,+1)$\\[3pt]
12 & 2 & $(x_1+\tfrac12,-x_2+\tfrac12,-x_3,x_5,x_4,-x_4-x_5-x_6,+1)$\\[3pt]
11 & 3 & $(-x_1,x_2+\tfrac12,-x_3+\tfrac12,x_6,-x_4-x_5-x_6,x_4,+1)$\\[3pt]
2 & 4 & $(-x_1+\tfrac12,-x_2,x_3+\tfrac12,-x_4-x_5-x_6,x_6,x_5,+1)$
\end{tabular}
\end{ruledtabular}
\end{table}

All eight remaining operations have threefold rotational parts.
Operations 3, 4, 5 and 7 add $2/3$ to each internal coordinate, giving
$\Delta\Sigma=8/3\equiv2/3$ modulo one; operations 6, 8, 9 and 10 add
$1/3$ to each, giving $\Delta\Sigma=4/3\equiv1/3$. For example,
operation 3 acts internally as
\[
 (x_4,x_5,x_6,x_7)\mapsto
 (x_4+\tfrac23,x_7+\tfrac23,x_5+\tfrac23,x_6+\tfrac23).
\]
It therefore maps the $\Sigma=0$ section to $\Sigma=2/3$, rather
than preserving it. The threefold operations are excluded automatically
by Eq.~\ref{eq:sm_section_test}. They remain operations of the extended
$3+4$D family, where they relate sections with different phase sums.

Time reversal combined with a half-period shift in all four internal
coordinates does preserve the section, because $\Delta\Sigma=2$.
After eliminating $x_7$, this operation becomes
\begin{equation}
 (x_1,x_2,x_3,x_4+\tfrac12,x_5+\tfrac12,x_6+\tfrac12,-1),
 \label{eq:sm_reduced_time_reversal}
\end{equation}
exactly the nonidentity centering entry in the ISODISTORT export.
Combining it with the four rows of
Table~\ref{tab:sm_four_arm_reduction} gives all eight magnetic operations
modulo lattice translations. These reproduce the orthorhombic MSSG
No.~19.3.146.7.m26.2, with basic space group $P2_12_12_1$, listed in
Table~\ref{tab:sm_lower_groups}. The comparison distinguishes the number
of active arms, the number of independent modulation coordinates, and
the symmetry retained at a fixed phase sum.

\subsection{Illustrations of the ideal lower-arm fields}

\begin{figure}[!ht]
\centering
\includegraphics[width=\textwidth]{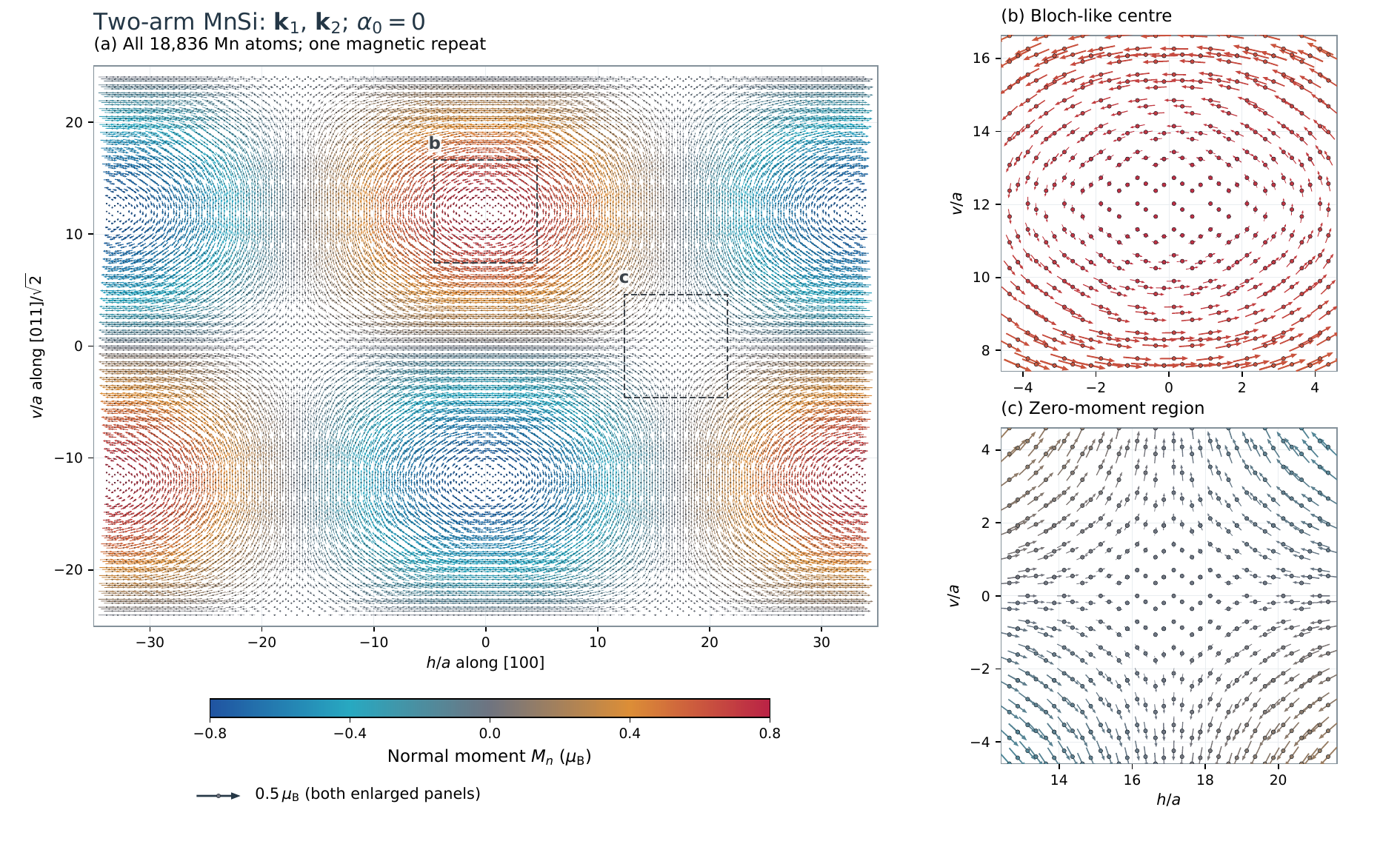}
\caption{Ideal two-arm ($3+2$D) field of Eq.~\ref{eq:sm_field2},
viewed along $\hat{\mathbf n}=[0\bar{1}1]/\sqrt2$, so that
$\mathbf k_1$ and $\mathbf k_2$ lie in the screen plane.
The horizontal and vertical directions are $[100]$ and $[011]$.
(a) All 18\,836 Mn atoms within one rectangular magnetic repeat,
$1/k$ by $1/(\sqrt2 k)$ in units of $a$, and a slab of thickness
$\sqrt2a$. This depth covers one crystallographic translation;
the continuous field is independent of depth in this direction.
The boxed regions are enlarged in (b) and (c), retaining all
477 and 513 Mn atoms, respectively. Panel (b) shows circulation
around a Bloch-like centre: the in-plane moment vanishes at the
centre, while the normal component remains finite. Panel (c)
shows the region around a zero of the full moment.
Atom positions lie at arrow midpoints. Arrow length represents the
projected moment; color denotes the moment component normal to the screen.
In (a), shaft width is proportional to the full moment magnitude.
In (b) and (c), both length and width scale with the projected moment,
giving identical slender arrow shapes and a common reference scale.}
\label{fig:sm_2k_atoms}
\end{figure}

\begin{figure}[!ht]
\centering
\includegraphics[width=\textwidth]{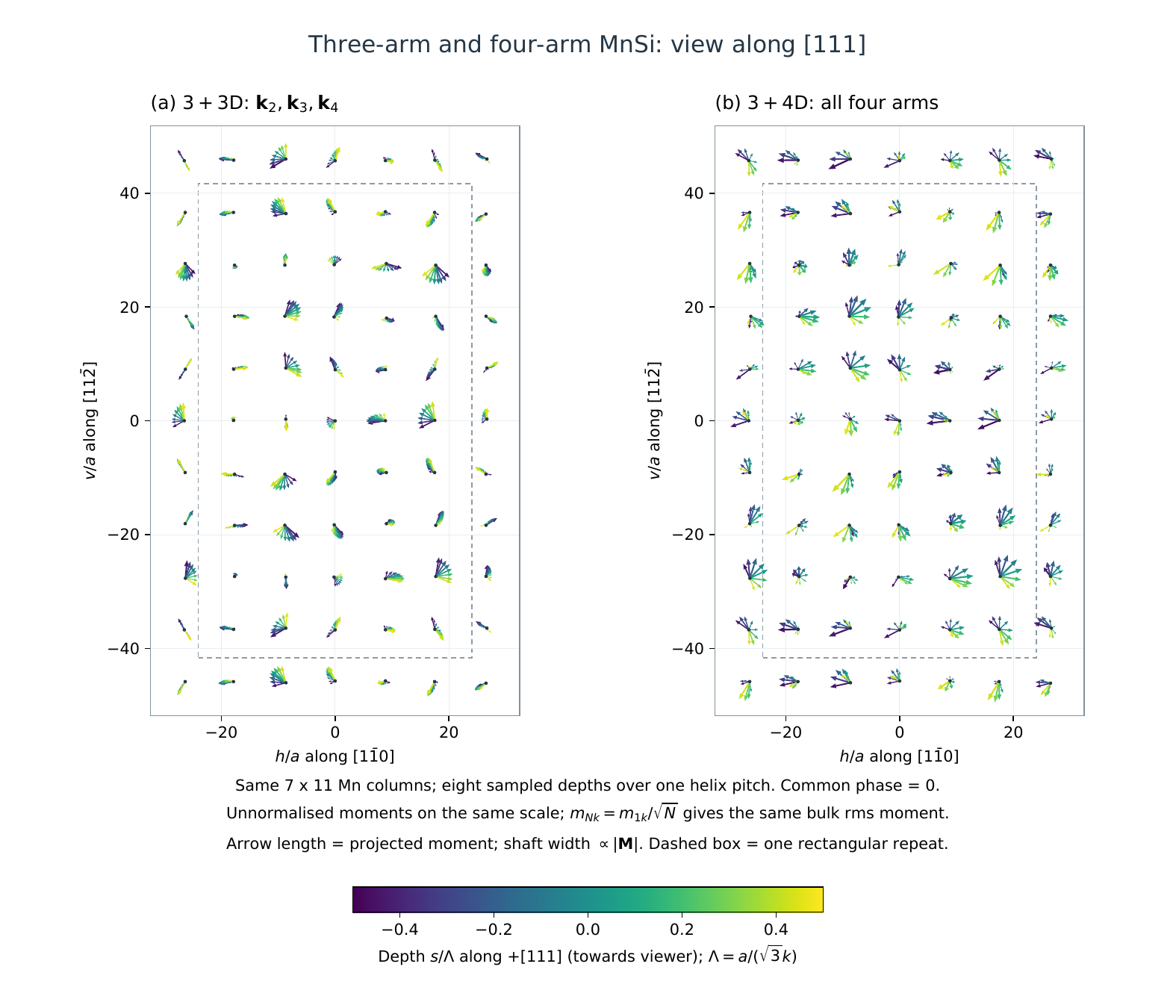}
\caption{Ideal (a) three-arm ($3+3$D) field with
$\mathbf k_2,\mathbf k_3,\mathbf k_4$ and (b) four-arm ($3+4$D) field,
viewed along $[111]$. Both panels show the same $7\times11$ Mn
columns and eight sampled atomic depths per column, spanning
$\Lambda=a/(\sqrt3 k)$ along $[111]$. Color encodes depth, with
positive depth toward the viewer. The horizontal and vertical
axes are along $[1\bar10]$ and $[11\bar2]$; dashed boxes mark a
rectangular transverse repeat. Arrow lengths and widths represent
projected and full moment magnitudes, respectively, on common
scales in the two panels. For the three inclined arms in (a), the
phase advances by only one third of a turn across this slab;
the full depth repeat is $3\Lambda$. Identical sampled atoms and
depths are used to isolate the change produced by adding the
fourth arm, with the stated mean-square-moment normalization.}
\label{fig:sm_3k_4k_columns}
\end{figure}

Figures~\ref{fig:sm_2k_atoms} and \ref{fig:sm_3k_4k_columns} illustrate
the fields above and their four-arm counterpart, using $k=0.0147$
and $\alpha_0=0$. The common normalization $m_{Nk}=m_{1k}/\sqrt N$,
with $m_{1k}=0.4371\,\mu_{\rm B}$, gives a bulk root-mean-square
moment of $0.5353\,\mu_{\rm B}$ in each construction. Local moments
are not normalized. The two-arm figure contains every Mn atom in
the stated slab, whereas the three-/four-arm comparison uses identical
sampled columns and depths for clarity. These figures illustrate
particular model configurations and do not identify which one is
realized in the sample.

\clearpage
